\documentclass[manuscript]{acmart}

\AtBeginDocument{%
  }

\setcopyright{acmlicensed}
\copyrightyear{2018}
\acmYear{2018}
\acmDOI{XXXXXXX.XXXXXXX}

\acmConference[Conference acronym 'XX]{Make sure to enter the correct
  conference title from your rights confirmation emai}{June 03--05,
  2018}{Woodstock, NY}
\acmISBN{978-1-4503-XXXX-X/18/06}

\usepackage{graphicx}
\begin{document}

\title{Towards Natural Language Environment: Understanding Seamless Natural-Language-Based Human-Multi-Robot Interactions}

\author{Ziyi Liu}
\authornote{Both authors contributed equally to this research.}
\affiliation{%
  \institution{School of Mechanical Engineering \\ Purdue University}
  \city{West Lafayette}
  \country{USA}}
\email{liu1362@purdue.edu}

\author{Xinyi Wang}
\authornotemark[1]
\affiliation{%
  \institution{Purdue University}
  \streetaddress{610 Purdue Mall}
  \city{West Lafayette}
  \state{IN}
  \country{USA}
  \postcode{47907}
}
\email{Wang6185@purdue.edu}

\author{SHAO-KANG HSIA}
\authornotemark[1]
\affiliation{%
  \institution{School of Mechanical Engineering \\ Purdue University}
  \streetaddress{610 Purdue Mall}
  \city{West Lafayette}
  \state{IN}
  \country{USA}
  \postcode{47907}
}
\email{shsia@purdue.edu}

\author{Chenfei Zhu}
\affiliation{%
  \institution{School of Mechanical Engineering \\ Purdue University}
  \streetaddress{610 Purdue Mall}
  \city{West Lafayette}
  \state{IN}
  \country{USA}
  \postcode{47907}
}
\email{zhu1237@purdue.edu}

\author{Zhengzhe Zhu}
\affiliation{%
  \institution{Elmore Family School of Electrical and Computer Engineering \\ Purdue University}
  \streetaddress{610 Purdue Mall}
  \city{West Lafayette}
  \state{IN}
  \country{USA}
  \postcode{47907}
}
\email{zhu714@purdue.edu}

\author{Xiyun Hu}
\affiliation{%
  \institution{School of Mechanical Engineering \\ Purdue University}
  \streetaddress{610 Purdue Mall}
  \city{West Lafayette}
  \state{IN}
  \country{USA}
  \postcode{47907}
}
\email{hu690@purdue.edu}

\author{Anastasia Kouvaras Ostrowski}
\orcid{0000-0001-8639-5135}
\affiliation{%
  \institution{School of Applied and Creative Computing \\ Purdue University}
  \city{West Lafayette}
  \country{USA}}
\email{akostrow@purdue.edu}

\author{Karthik Ramani}
\orcid{0000-0001-8639-5135}
\affiliation{%
  \institution{School of Mechanical Engineering \\ Purdue University}
  \city{West Lafayette}
  \country{USA}}
\affiliation{%
  \institution{Elmore Family School of Electrical and Computer Engineering \\ Purdue University}
  \city{West Lafayette}
  \country{USA}}
\email{ramani@purdue.edu}


\renewcommand{\shortauthors}{Trovato et al.}

\begin{abstract}
As multiple robots are expected to coexist in future households, natural language is increasingly envisioned as a primary medium for human–robot and robot–robot communication. This paper introduces the concept of a Natural Language Environment (NLE), defined as an interaction space in which humans and multiple heterogeneous robots coordinate primarily through natural language. 

Rather than proposing a deployable system, this work aims to explore the design space of such environments. We first synthesize prior work on language-based human–robot interaction to derive a preliminary design space for NLEs. We then conduct a role-playing study in virtual reality to investigate how people conceptualize, negotiate, and coordinate human–multi-robot interactions within this imagined environment. 

Based on qualitative and quantitative analysis, we refine the preliminary design space and derive design implications that highlight key tensions and opportunities around task coordination dominance, robot autonomy, and robot personality in Natural Language Environments.
\end{abstract}



\begin{teaserfigure}
  \includegraphics[width=\textwidth]{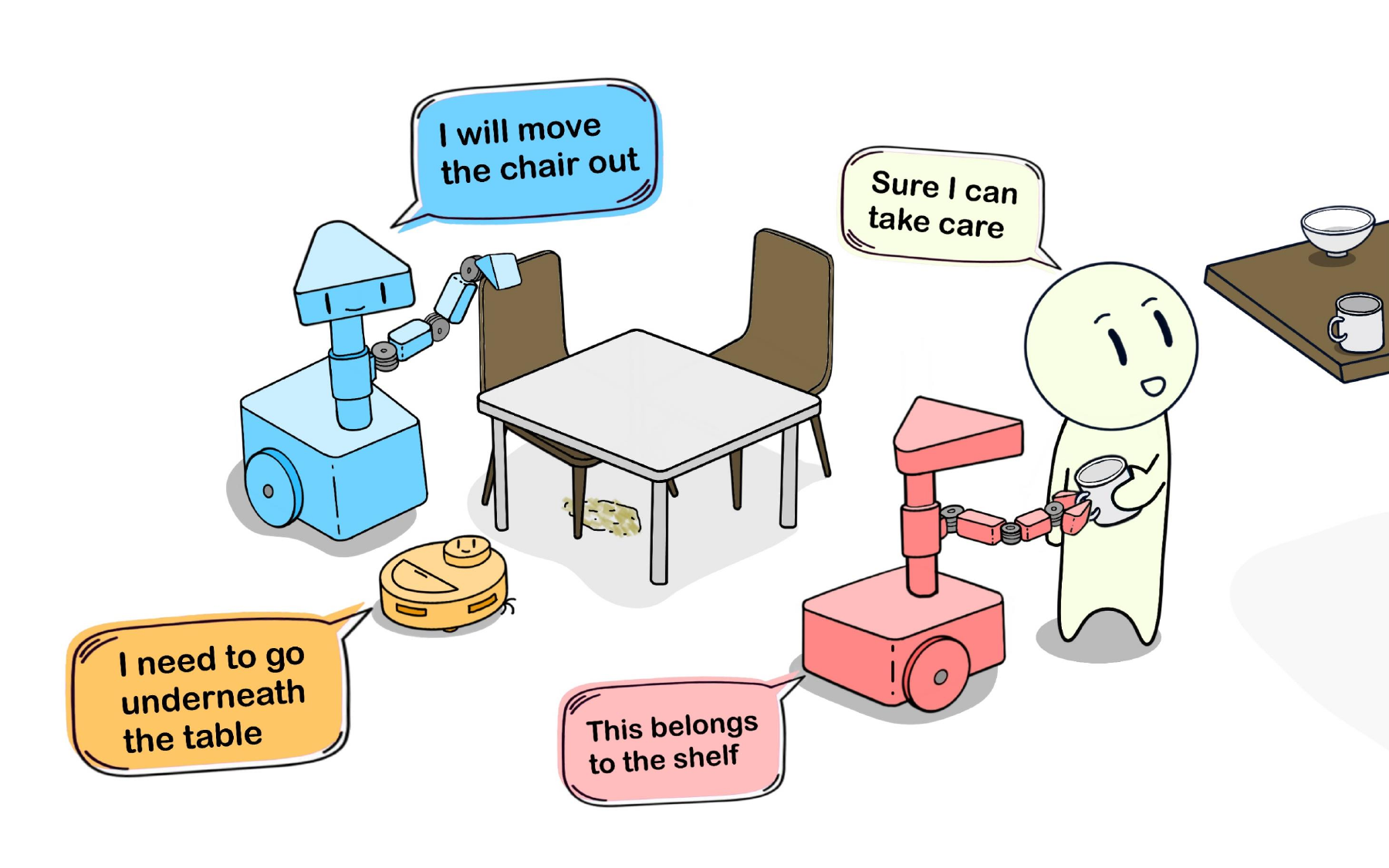}
  \caption{The concept of Natural Language Environment (NLE) in Human-Robot Interaction (HRI): a shared space where humans and multiple robots with varying physical capabilities can seamlessly interact and collaborate using natural language.}
  \Description{Replace in the future}
  \label{fig:teaser}
\end{teaserfigure}

\received{20 February 2007}
\received[revised]{12 March 2009}
\received[accepted]{5 June 2009}

\maketitle

\section{Introduction}
As robots advance toward becoming sophisticated personal assistants, their integration into ordinary households is accelerating. This shift reflects a broader technological trend in which domestic robots increasingly handle tasks that enhance convenience, safety, and comfort in daily life \cite{AHumanoidSocialRobotBased, DesigningParent-child-robot, Largelanguagemodelsforhuman, ApplicationofRoboticstoDomestic}. From cleaning robots \cite{ApplicationofRoboticstoDomestic, SmartCleaner:ANewAutonomous, AutonomousSelf-ReconfigurableFloor, AHumanSupportRobotforthe, sTetro-D} to assistive robots \cite{amazonastro2024,meetobi2024, mobilemanipulation, PaLME, bostondynamics2024}, educational robots \cite{DesigningParent-child-robot, RobocampatHome:Exploring, Robottutorandpupils’educational}, and social robots \cite{AMultimodalEmotionalHuman–Robot, AHumanoidSocialRobotBased, Mini:ANewSocialRobot}, these systems are no longer futuristic concepts but are gradually becoming part of everyday life. As domestic robots continue to diversify and become more affordable and capable, it is foreseeable that a single household will be equipped with multiple robots. Such multi-robot household scenarios introduce new opportunities and challenges for human–robot interaction (HRI), particularly with respect to communication modalities, task coordination, autonomy levels, personalization, and robot personalities \cite{AMultirobotSysteminanAssisted, ConversationalLanguageModelsfor, designofahomemulti, HowManyRobotsDoYouWant?, ASmartHomeBased}.

Humans naturally gravitate toward intuitive communication methods, with natural language being one of the most fundamental forms of interaction \cite{RobotsThatUseLanguage, socialrobot, socialrobotforlongterm}. This preference is reflected in the widespread adoption of voice-based assistants such as Amazon Alexa \cite{alexa}, Google Assistant \cite{GoogleAssistant}, and Apple’s Siri \cite{AppleSiri}. More recently, the emergence of multimodal large language models (LLMs), such as GPT-4 \cite{openai2024gpt4technicalreport} and PaLM \cite{palm2}, has further strengthened this trend by enhancing robots’ natural language processing and in-context scene understanding capabilities. These models enable robots to interpret and respond to their surroundings in more flexible and natural ways \cite{huang2024rekepspatiotemporalreasoningrelational, WhatsonYourMind, EnablingRobotstoUnderstand, correctingrobotplansnatural}.

Together, these developments suggest a shift toward communication between humans and robots that is increasingly compact, implicit, flexible, and natural. As these technologies continue to evolve, it is conceivable that robots will eventually achieve human-level capabilities in natural language processing, contextual awareness, and scene understanding, enabling seamless natural-language-based communication with humans, and potentially among robots themselves. Motivated by this vision, we define the concept of a \textbf{Natural Language Environment (NLE)} in Human–Robot Interaction (HRI): a shared environment in which humans and multiple robots with varying physical capabilities interact and collaborate primarily through natural language.

Although extensive research has explored language-based HRI \cite{UnderstandingLarge-LanguageModelLLM-powered, TheConversationistheCommand, LanguageModelsforHuman-Robot, CreatingPersonalizedVerbalHuman-Robot, InteractiveLanguage, NototheRight, TidyBot, Text2Motion, ExplainingAutonomy, opensource, Reshapingrobot, correctingrobotplansnatural, bringinganatural, RobotsThatUseLanguage, RecognisingFlexibleIntent, Language-ConditionedImitation, WhatsonYourMind, ExploringtheImpactofExplanation, ExplainableAIforRobotFailure}, comparatively little attention has been paid to scenarios in which humans and multiple robots with heterogeneous capabilities communicate and collaborate using natural language—a setting that remains largely unexplored within NLEs. Such environments pose a range of challenges, including appropriate robot autonomy levels, coordination and dominance in collaborative tasks, suitable communication strategies, and differences in robots’ physical capabilities. These challenges underscore the need to study NLEs systematically, develop an initial design space, and derive design insights for human–robot interaction within such environments.

To investigate these challenges and gain insights into effective communication between humans and multiple robots in an NLE, we conducted a user study that simulates such interactions. Participants assumed the roles of humans and robots and interacted exclusively through natural language within a virtual reality (VR) environment. This role-playing setup enabled us to observe interaction dynamics and explore potential approaches to addressing the challenges outlined above.

Through this work, we make the following contributions:
\begin{itemize}
    \item We introduce the concept of a Natural Language Environment (NLE) in Human–Robot Interaction and propose a preliminary design space grounded in seamless, human-like communication between humans and robots.
    \item We present a VR-based user study that employs role-playing to simulate and examine interactions between humans and multiple domestic robots within an NLE.
    \item We derive a set of design implications from our study to refine and inform the proposed design space for NLEs.
\end{itemize}

\section{Related Work}

\subsection{Trends in domestic robot}
 
Recent advances in computer technologies have led to the development of more intelligent and versatile robots for domestic use. With improvements in perception, recognition, decision-making, path-planning, and communication, domestic robots are increasingly capable of assisting with household duties such as cleaning, educating children, monitoring the environment, and providing companionship \cite{Largelanguagemodelsforhuman}. domestic robots are designed with diverse capabilities based on their intended tasks. Among them, cleaning robots and assistive robots are the most common.


\textit{Cleaning robots} are equipped with advanced sensing, path-planning, and navigation technologies to aid in domestic cleaning. Early cleaning robots were limited to basic floor cleaning with elementary navigation and obstacle avoidance capabilities \cite{Autonomousvacuumcleaner}. Over time, advances in sensor technology, mapping algorithms, and machine learning have significantly enhanced the capabilities of cleaning robots, allowing them to navigate and clean complex home environments \cite{EvaluatingtheRoomba}. Modern cleaning robots can tackle tasks beyond flat floors, including windows \cite{ASurveyonTechniquesandApplications, ASwitchableUnmanned}, walls \cite{EEG-ControlledWall-Crawling}, tables \cite{TableCleaningTask}, stairs \cite{sTetro-D} and even door handles \cite{AHumanSupportRobotforthe}. 

\textit{Assistive robots} are designed to help manipulate objects for various tasks at home. Recent progress in object detection, path planning, and navigation have made these robots increasingly versatile.
Stationary assistive robots \cite{lotenableddualarm} are typically mounted on tables or floors to assist with object manipulation, while mobile assistive robots \cite{mobilemanipulation} are enhanced with mobility and navigation ability. The integration of natural language processing (NLP) has further enhanced their capabilities. Google's PaLM-E \cite{PaLME} uses a language model to process raw sensor data for effective task execution, and Yenamandra et al. \cite{homerobot} developed a compliant robot that performs open-vocabulary object manipulation. These advancements enable assistive robots to understand human commands and and more adaptively interact with their environment.

Other domestic robots like social robots \cite{AMultimodalEmotionalHuman–Robot, AHumanoidSocialRobotBased, Mini:ANewSocialRobot}, educational robots \cite{DesigningParent-child-robot, RobocampatHome:Exploring, Robottutorandpupils’educational}, and health care robots \cite{ENRICHME:Perception, IntegratingSocialAssistiveRobots, Home-BasedRehabilitationof} are also gaining popularity. As robotic capabilities expand and costs decrease, it’s easy to envision a future where households are equipped with multiple robots. They, each specializing in different tasks, can handle a variety of domestic tasks at users' command. With the ubiquity of domestic robots, users face the challenge of efficiently and intuitively communicating their intent for robots to carry out tasks effectively \cite{Asurveyofmultiagent, ConversationalLanguageModelsfor, RoCo:DialecticMulti-Robo}. The challenge intensifies especially when specific tasks require coordination from human-to-robot or robot-to-robot to complete.

\subsection{Natural Language in Human-Robot Interaction}

People use language every day to direct behavior, ask and answer questions, provide information, and ask for help. Compared to traditional keyboard-and-mouse or touch-screen interfaces, language-based commands require minimal user training and allow the expression of a variety of complex tasks, making them preferred by humans, especially in household scenarios where robot experts are absent. \cite{RobotsThatUseLanguage}. Communicating with robots using natural language has been studied for a long time.

Early attempts utilized predefined \cite{Mobilerobotprogramming, Spatiallanguagefor} or structured language commands \cite{walkthetalk, towardsunderstanding, understandingnaturallanguagecommands} to control robot behaviors.
Due to the limited perception and reasoning abilities of robots at the time, interactions were largely one-way, with robots merely following commands without engaging in more conversational exchanges. 
Kollar et al. \cite{towardsunderstanding} presented a system that follows natural language directions by extracting a sequence
of spatial description clauses from the linguistic input and then infers the most probable path through the environment.
Lauria et al. \cite{Mobilerobotprogramming} provided the robot with a set of primitive procedures derived from a corpus of route instructions to enable unconstrained language instructions on navigation. 
While these methods could parse and execute language commands, their limited scope of language expressions constrained the naturalness and flexibility of interactions. This limitation prevented them from supporting high-level abstract instructions with implicit information, which is typical in human language conversations \cite{groundedlanguage}.

To enhance language interactions, subsequent approaches moved beyond rigid language structures, enabling the parsing of high-level and abstract commands and breaking them down into executable robot actions \cite{EnablingRobotstoUnderstand, APersistentSpatialSemantic, learningtoparse, naturallanguagedecomposition}. This shift allowed for more natural and flexible communications between users and robots. 
Chen et al. \cite{EnablingRobotstoUnderstand} introduced a new method to perform abstract natural language instructions by automatically filling in information missing from the instruction using environmental context and a new commonsense reasoning approach.
 Blukis et al. \cite{APersistentSpatialSemantic} proposed a persistent spatial semantic representation method to enable building an agent that performs hierarchical reasoning to effectively execute long-term tasks from a high-level language instruction. 
However, despite improvements in interaction flexibility and the robots' ability to understand the underlying intents in high-level language commands, these methods still struggle with fully grasping and interpreting implicit information in language, resulting in a lack of context awareness and continuity in robots' responses \cite{AReviewofNaturalLanguage}.

The emergence of LLMs has significantly transformed human-robot interactions. LLMs surpass traditional language processing models in reasoning, scene understanding, and contextual awareness, leading to better extraction of implicit information and more natural dialogue generation \cite{palm2, openai2024gpt4technicalreport}. 
Robots equipped with LLMs \cite{LaMI, GenerativeExpressive, LM-Nav, CreatingPersonalizedVerbalHuman-Robot, LanguageModelsforHuman-Robot, TheConversationistheCommand, Text2Motion} can more accurately understand user intents, interpret complex and high-level instructions, and provide natural, coherent, and context-aware responses. This advancement enables more sophisticated and meaningful bidirectional communication between humans and robots. 
Billing et al. \cite{LanguageModelsforHuman-Robot} first integrated the GPT-3 language model with the Aldebaran Pepper and Nao robots to achieve an open verbal dialogue with them.
Onorati et al. \cite{CreatingPersonalizedVerbalHuman-Robot} presented a social robot application for conducting personalized conversations using social media data for the user and large-language models to build the dialogue.
Lin et al. \cite{Text2Motion} used feasibility heuristics encoded in Q-functions of a library of skills to guide task planning with LLMs.
Nwankwo and Rueckert \cite{TheConversationistheCommand} leveraged the LLMs to decode the high-level natural language
instructions and abstract them into precise robot actionable commands or queries and utiliszd a vision-language model to provide a visual, and semantic understanding of the robot’s task environment.

As natural language processing technology continues to evolve, future human-robot interactions may become as seamless and natural as human-to-human communication, with robots achieving human-level language capabilities.

\subsection{Human Multi-Robot Interaction}

As the prevalence and diversity of domestic robots and home automation increases, it will become common for homes to be equipped with various robots to assist with different tasks, where each robot does some category of tasks well. This situation presents new challenges for human-robot interaction, specifically in how people can communicate with multiple robots in a natural, flexible, and efficient way. Previous research has explored different approaches to address this challenge. 
Hunt et al. \cite{ConversationalLanguageModelsfor} proposed a dialogue-based approach, allowing robots with different abilities to collaboratively plan solutions through peer-to-peer and human-robot discussions.
Kannan et al. \cite{SMART-LLM:SmartMulti-Agent} utilized LLMs to convert high-level task instructions into multi-robot task plans.
Lestingi et al. \cite{Formalmodelingandverification} developed a framework for human-robot interaction scenarios that estimate the probability of mission success, addressing the complexity of multi-agent systems and the unpredictability of human behavior.  
Zhao et al. \cite{RoCo:DialecticMulti-Robo} employed LLMs for collaborative task strategy reasoning and waypoint path generation in multi-robot environments.
Wang et al. \cite{PushThatThereTabletop} introduced a multi-modal object-level method for autonomous and collective object manipulation by a multi-robot system.
Barber et al. \cite{AMultirobotSysteminanAssisted} designed a multi-mobile robot system integrated into an automated home to monitor the well-being of the elderly, assist with household tasks, and provide emotional coaching.
Zhang et al. \cite{ASmartHomeBased} proposed a smart home integrating a heterogeneous robot system and distributed sensor networks to support a wide range of activities for elderly care.

While prior research has proposed different methods to achieve communication between humans and multi-robots, there has not been a work studying the aspects that need to be considered when designing applications for human multi-robot interaction with natural language in homes. To this end, we propose the concept of a Natural Language Environment, where natural language is the main medium of communication between humans and robots as well as between robots. Through literature review and user study, we proposed a design space tailored for the envisioned environment.

\section{Natural Language Environment}

We define a Natural Language Environment (NLE) in Human–Robot Interaction (HRI) as a conceptual interaction space in which humans and multiple robots with heterogeneous physical capabilities coordinate primarily through natural language. In this work, NLE is not treated as an already-realized technical system, but as a design vision that allows us to reason about future interaction paradigms and their implications.

Rather than assuming fully mature technology, we treat NLE as a speculative but grounded design context. Our goal is to use this concept to explore how people imagine, negotiate, and design interactions in such environments, given current trends in robotics and large language models. This allows us to study potential interaction patterns, tensions, and expectations before such environments become technically feasible.

\subsection{Assumptions for Natural Language Environment}
We made the following assumptions as we defining a Natural Language Environment:

\textbf{Assumption 1: Human-Multi-Robot Collaboration is Essential in a NLE.} The increasing prevalence and diversity of domestic robots result in the coexistence of multiple robots alongside humans in domestic environments. Due to the various capabilities across characters, no single robot or human is supposed to handle all household tasks alone, effective teamwork that leverages the strengths of different robots and humans becomes crucial.

\textbf{Assumption 2: Robots have human-level perception capabilities in a NLE.} Progress in computer vision and artificial intelligence have granted robots some degree of perceptual abilities, enabling them to understand scenes, recognize objects, and adapt to various contexts. While current robots' perception is not yet on par with humans, future developments are expected to equip robots in the NLE with human-level perception capabilities.

\textbf{Assumption 3: Robots have human-level natural language capabilities in a NLE.} Enhancing natural language capabilities is a key focus in robotics. Advanced natural language skills are crucial for robots to effectively collaborate with and assist humans in household tasks. Future robots are expected to have human-level natural language capabilities, allowing them to interact seamlessly with humans and other robots and assist with various household tasks.

\subsection{Preliminary Design Space in Natural Language Environment}
NLE involves natural, seamless, and varied natural language communication between humans and multiple robots and among multiple robots themselves, which makes the design space for NLE much more sophisticated than a design space for human-robot interaction with limited language communications. However, since the main communications within the NLE involve human-robot and robot-robot, the design space of NLE shares some similarities with the design space of previous natural-language-based HRI works. To systematically explore the design space of NLE, we gathered and summarized the design spaces from those works. Grounded by the characteristics of the robots in a NLE, and our review of the related literature, we propose a preliminary design space for NLE, which addresses three key dimensions: Dominance of task coordination, Level of Autonomy, and Personality.

\subsubsection{Dominance of Task Coordination}
The dominance of task coordination refers to the lead of task coordination in human-robot collaborations. In the context of human-robot communication in natural language, the dominance of task coordination often refers to the leader who leads the discussion and assigns tasks. Due to the differences in expertise of humans and robots, when the two work together, the dominance often changes as per the task. Based on who holds the dominance, we categorize the dominance of task coordination into three types.

\textbf{Human-dominant}. In human-dominant task coordination, humans lead the tasks, assigning roles to robots and making key decisions. Human-dominant task coordination gives humans full control of the task process and robot actions. It fosters trust because humans directly manage how robots contribute to the task, and is often preferred in tasks where human oversight is non-negotiable \cite{PreferredInteractionStyles, InfluencingLeadingandFollowing, DecisionMakingAuthority}. On the other hand, human-dominant task coordination can be inefficient and may increase the cognitive load of humans, as humans need to instruct robots continuously \cite{Theeffectofcognitive}.

\textbf{Robot-dominant}. In robot-dominant task coordination, the robot takes the lead in making decisions, while the human acts as a follower. Robot-dominant task coordination can increase the efficiency of collaborative work, as all the decisions are made towards a better and more efficient completion of tasks \cite{DecisionMakingAuthority}. It works well for tasks where the capabilities of the robots surpass that of humans. However, robot-dominant task coordination may cause trust issues when the robot's decisions are unclear or not aligned with human expectations, as it reduces the sense of control of humans. It is essential to ensure transparency and allow human's intervention during a robot's performance of a task to prevent discomfort or disengagement \cite{RobotleadershipInvestigating, TheWorldisnotEnough, Collaboratingeyetoeye}. 

\textbf{Adaptive-dominant}. Adaptive dominance allows the leadership to shift between humans and robots based on the task performed. Robots can take the lead when efficient automation is needed but seamlessly defer to human control when tasks become more complex or require human decisions. Adaptive dominance boosts human-robot task coordination as the human and robots need to discuss and assign the leadership jointly and each party utilizes its own advantages in leading the task \cite{Adaptiveassistiverobotics}. Adaptive dominance achieves a balance between task efficiency, human trust, and transparency, while achieving seamless leadership shifts between multiple characters may require sophisticated communication and task coordination \cite{AdaptiveLeaderFollower, Humanrobotmutual}.

\subsubsection{Level of Autonomy}\label{autonomy}

The level of autonomy is a crucial factor in human-robot interaction, significantly influencing how robots interpret and act on human commands and determine their communication with humans. It plays an important role in shaping human-robot collaboration, task performance, and human trust \cite{Towardaframeworkforlevels}. Based on prior research and our definition and assumptions for NLE, we categorize robot autonomy into three levels:

\textbf{Teleoperation Autonomous}: This lowest level of autonomy involves entire control over the robot's actions by humans or other robots. 
At this autonomy level, the robot acts merely as an executor of the intentions of other characters, similar to traditional teleoperation systems. 
While this level allows for high precision and reliability due to complete control over the robot's actions, it can be less efficient because of the need for continuous communication and manipulation \cite{Teleroboticsautomation, HumanPerformanceIssues}.

\textbf{Semi Autonomous}:
This intermediate level features a collaborative decision-making process between robots and humans or other robots. Semi autonomy integrates the robot's capabilities with input from other characters to perform tasks. Robots at this level can make decisions independently but seek additional information or assistance when dealing with complex or uncertain scenarios. This autonomy level exemplifies human-robot and robot-robot collaboration, leveraging both autonomous capabilities and the benefits of joint decision-making \cite{HumanRobotMutualAdaptation}.

Based on the information requested, we categorized Semi autonomy into two classes, similar to \cite{ataxonomyoffactors}:

\begin{itemize}
    \item Cognitive Autonomous: Robots can fully understand contexts and make cognitive decisions independently but may seek physical assistance from other characters to complete tasks.
    \item Physical Autonomous: Robots can perform physical actions independently but may request information from others to better understand the context.
\end{itemize}

\textbf{Full Autonomous}:
The highest level of autonomy allows robots to operate independently with minimal intervention from humans or other robots. Robots with full autonomy exhibit advanced decision-making abilities and environmental adaptability, enabling them to handle various complex situations on their own. While this level can greatly enhance efficiency \cite{thetransferofhuman}, it may also result in challenges related to explainability and control \cite{ASurveyonTrustinAutonomous, TrustAwareDecisionMaking}.

It is notable that NLE features diverse robots, multilateral human-robots interactions, and natural language communication, which makes autonomy in NLE different from general robot autonomy, as different types of robots may have different autonomy levels.

\subsubsection{Personality}
The personality of a robot plays a crucial role in determining how the robot executes tasks and communicates with humans or other robots \cite{PersonalityintheHumanRobot, TowardsanIntegrative}. The personality of a robot is often categorized into two aspects, sociality and anthropomorphism \cite{ThreeResponsesto}.  

\textbf{Anthropomorphism} refers to robots exhibiting human-like characteristics, such as emotions, facial expressions, or conversational tone \cite{Anthropomorphisminhumanrobot}. Some degree of anthropomorphism, such as mimicking a human's smile when collaborating to express acknowledgment to the human, has been shown helpful in increasing human trust and the likeability of robots. However, excessive human-like behavior can sometimes raise unrealistic expectations or cause discomfort, especially when the robot's capabilities do not match its appearance or behavior \cite{AnthropomorphismOpportunities}.

\textbf{Sociability} represents the robot’s ability to engage in social behaviors, including collaboration, appropriate behavior, and understanding social cues \cite{Asurveyofsocially}. Sociability can enhance a robot's ability to collaborate with humans and other robots, as well as gain reliance and trust from them \cite{Humanoidrobotsasa}. Sociality is often essential for successful and efficient human-robot collaboration, especially in the context of human-robot communication is achieved by natural language \cite{Towardsociablerobots}.

In NLE, humans work with multiple robots with different capabilities. The diversity of robot structures and capabilities in NLE provides a basis for assigning different personalities to different robots, as robots with different features may be expected to play different roles in a collaboration.

\section{Method}
This section describes the methodological framework used to investigate interaction design in Natural Language Environments. Rather than evaluating a deployed multi-robot system, we adopt a design research approach that uses role-playing in virtual reality to simulate key properties of NLEs. This approach allows us to examine how participants conceptualize, negotiate, and coordinate interactions in environments characterized by language-based multi-robot collaboration, while abstracting away from current technical limitations of robotic systems.

\subsection{Role-playing Methodology in Virtual Reality}

In order to understand the interactions between humans and domestic robots in a Natural Language Environment, we proposed a study to observe how humans collaborate with multiple domestic robots on a common house cleaning task in an ordinary household setting.  As robots have not yet reached the human level of language and perception capabilities as in our assumptions, we adopted a role-playing methodology in our study to imitate robots with such capabilities. Role-playing methodology is a participatory design technique that facilitates the exploration of user experiences and system interactions through enacted scenarios \cite{roleplay1, roleplay2}. This approach enables researchers to elicit rich, contextual data by simulating real-world usage situations, thereby enhancing understanding of user needs and potential design solutions\cite{roleplay1}. In our study, participants were requested to assume the roles of various domestic robots and interact with participants who represented actual humans.

To observe more realistic interactions between humans and multiple robots in a household setting, we utilized Virtual Reality (VR) as the medium of our study. VR technology has been widely adopted in the fields of education, healthcare, architecture, and tourism \cite{VRoverview}, given its uniqueness in providing highly immersive experiences. Jackson et al. identify four distinct dimensions of immersive experience in VR \cite{immersiveVR}. These dimensions include \textit{representational immersion}, which aligns with the concept of psychological presence, and \textit{participatory immersion}, which relates to the interactive elements of the VR experience. Additionally, \textit{affective immersion} pertains to the emotional connection one develops with the experience, while \textit{narrative immersion} refers to the engagement and involvement in the unfolding events within the VR environment \cite{immersiveVR}. This immersive experience allows participants to quickly adapt to their assigned roles in the simulated scenario, potentially improving the performance of the role-playing methodology. 

\subsection{VR Interface}
\begin{figure}[H]
  \includegraphics[width=\textwidth]{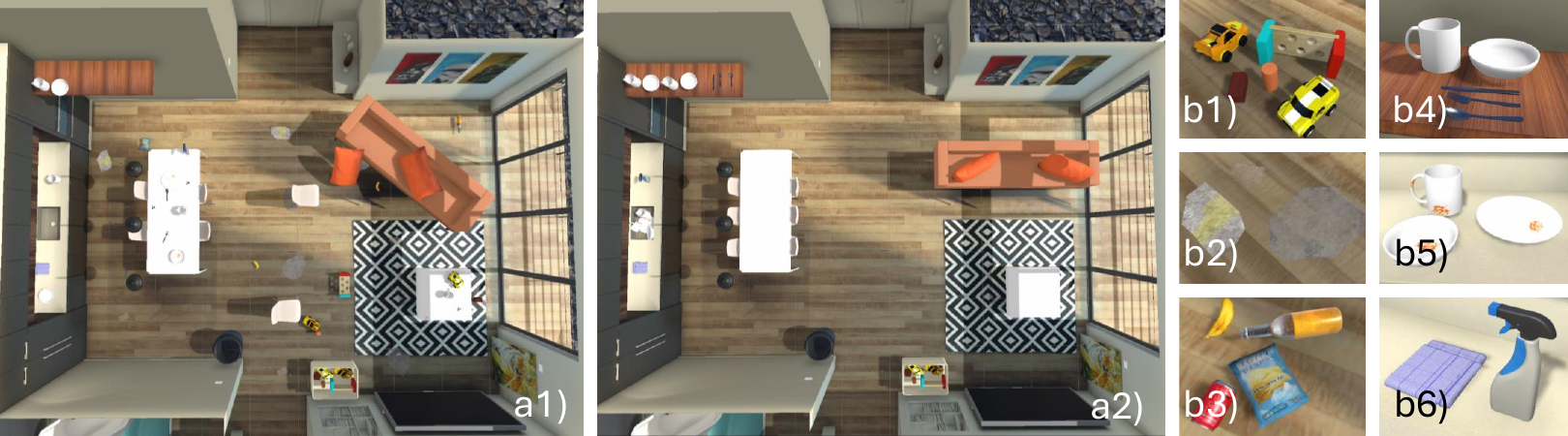}
  \caption{Bird's-eye view of the apartment setting and interactable objects. a1) The initial setting before house cleaning. a2) An example of house setting after house cleaning. b1) Toys. b2) Dust. b3) Trash. b4) Clean tableware. b5) Dirty tableware. b6) Towel and cleaner.}
   \Description{Replace in the future}
  \label{fig:room}
\end{figure}

\begin{figure}[H]
  \includegraphics[width=\textwidth]{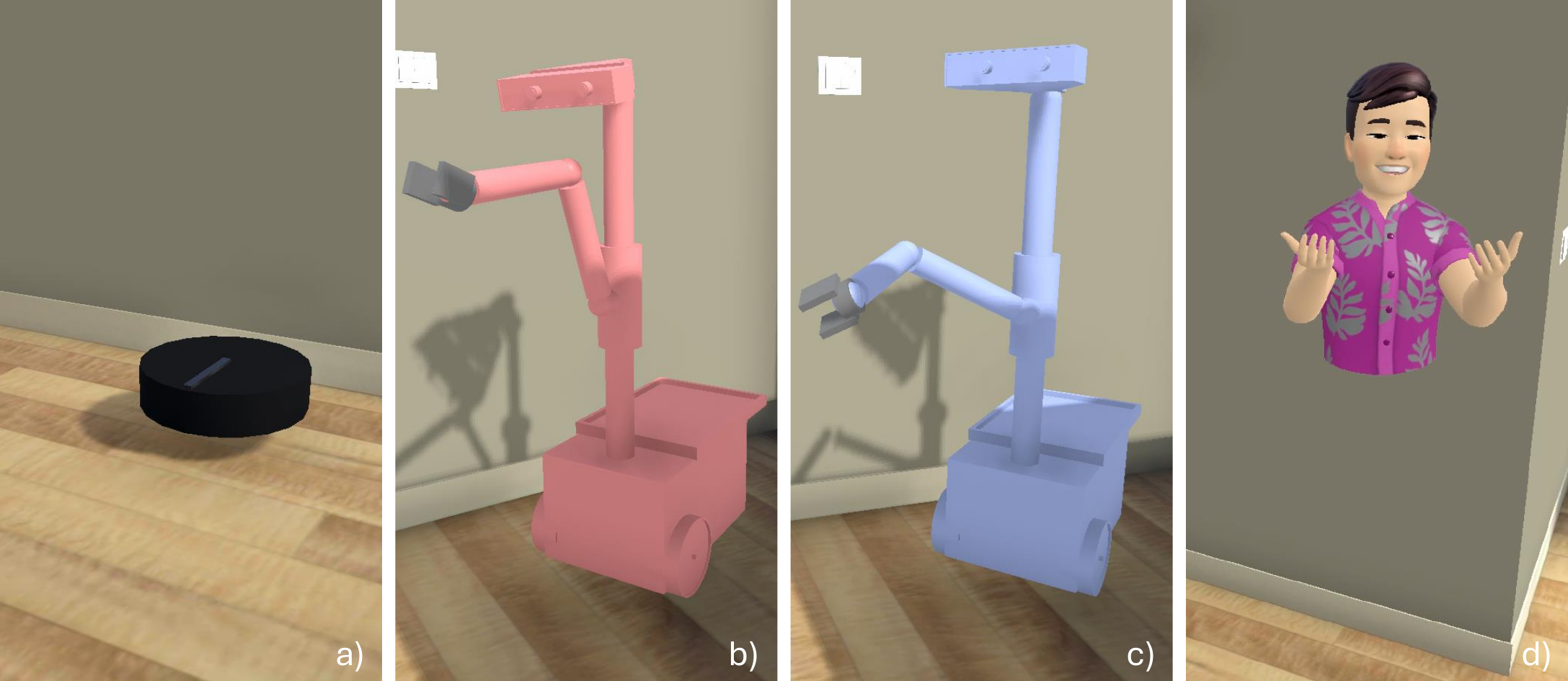}
  \caption{The four characters in the game: (a) the cleaning robot, (b) the first assistive robot, (c) the second assistive robot, and (d) the human.}
   \Description{Replace in the future}
  \label{fig:characters}
\end{figure}

In the virtual scene, we created an apartment with a living room, dining room, kitchen, bathroom, and bedroom (Figure \ref{fig:room} a), where four players work collaboratively on ordinary house cleaning tasks, such as cleaning and organizing. The scene includes four characters: one cleaning robot, two assistance robots, and one human, which are shown in Figure \ref{fig:characters}.

The cleaning robot, designed specifically for vacuuming, is included due to its wide acceptance in households. By incorporating this type of robot, we can observe how this commonly used robot interacts with others in a Natural Language Environment. 

\begin{figure}[H]
  \includegraphics[width=\textwidth]{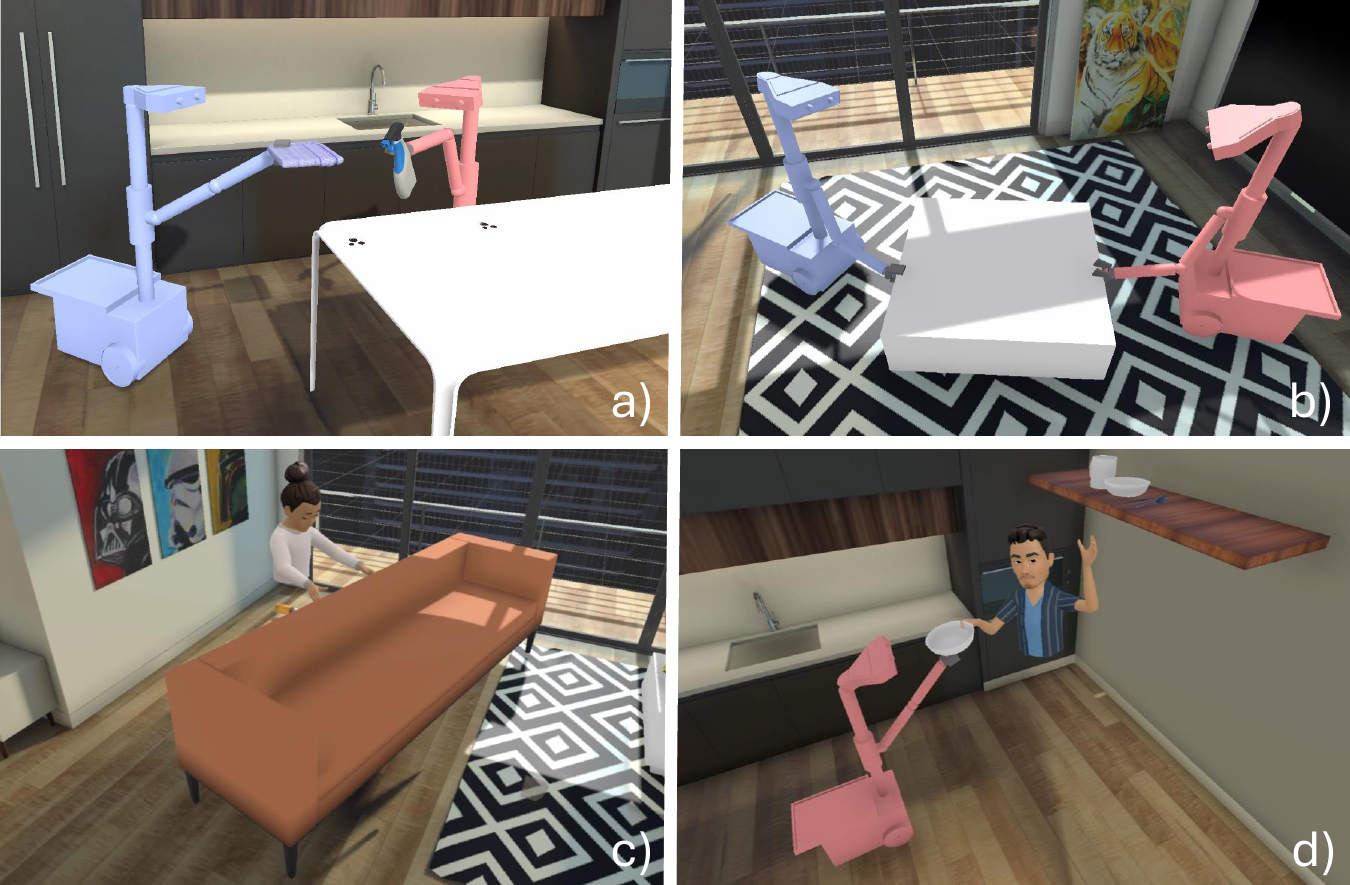}
  \caption{Some interactions between the characters. In (a), the two assistive robots work together to remove stains on a table. One robot applies cleaner on stains, while the other one wipes them off using a towel. (b) The two assistive robots adjust the coffee table collaboratively. (c) The human adjusts the heavy sofa by himself. (d)The red robot can not reach the high shelf and ask the human for help to put a clean bowl on the shelf.}
  \Description{Replace in the future}
  \label{fig:interactions}
\end{figure}

The two assistive robots are designed referred to the Google PaLM-E robot\cite{PaLME}, each featuring a mobile base and a robotic arm. This configuration represents the trend of versatile robotic systems capable of performing sophisticated tasks. Their capabilities include carrying objects, and interacting with various household items, making them ideal for studying coordination and collaboration in physical task execution. Figure \ref{fig:interactions} a) shows two assistive robots performing a house-cleaning task.

The virtual apartment consists of a set of interactable objects including dust, stains, trash, tableware, toys, and furnitures. Some of the small objects are shown in Figure \ref{fig:room} b). The physical properties of those virtual objects are inherited from their real-world counterparts. For example, based on the weight of an object, the assistive robot can move the chair independently, move the table with the assistance of another robot or a person, as shown in Figure \ref{fig:interactions} b), but can not move the sofa, which can only be moved by the human (Figure \ref{fig:interactions} c). As an additional example, the stains cannot be wiped out until the cleaner has been applied.These subtle elements create a powerful sense of immersion and emphasize the need for cooperation and communication among the robots and the human.

\subsubsection{Robot Interface}

\begin{figure}[H]
  \includegraphics[width=\textwidth]{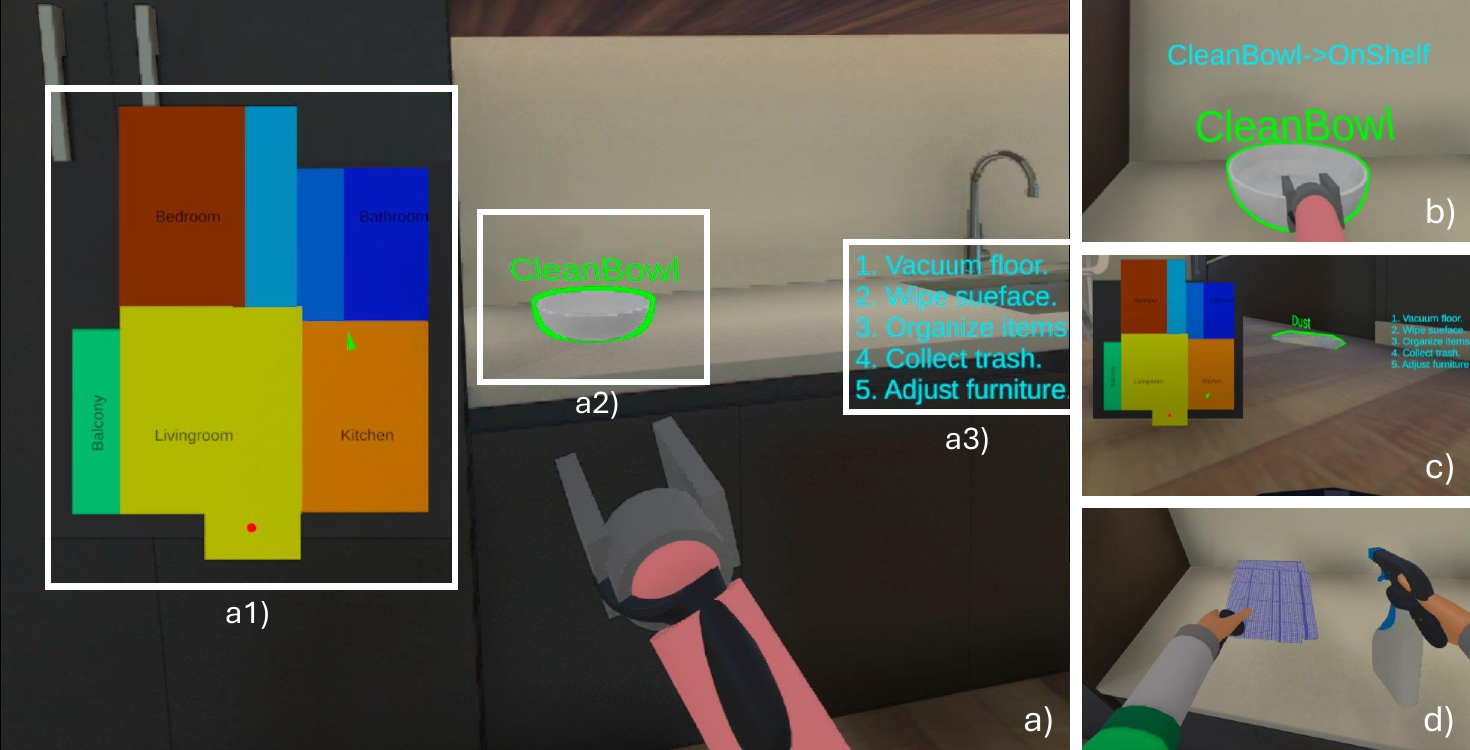}
  \caption{The VR views for the assistive robots, the cleaning robot, and the human. In the assistive robot view of (a), (a1) a SLAM map, (a2) an object detection simulation, and (a3) a to-do list are provided as generated by LLM. A hint is displayed when the assistive robot grabs an object, as shown in (b). (c) shows the cleaning robot view, where only dust is highlighted and labeled. The human view is depicted in (d), in which the human grabs two objects with both hands.}
  \Description{Replace in the future}
  \label{fig:interfaces}
\end{figure}

\textbf{\textit{Assistive robot.}}
Throughout the game, players controlling the robots remain seated and use the joystick on the left controller to navigate within the apartment. To enhance the player's environmental perception, a map, shown in Figure \ref{fig:interfaces} a1). is displayed at the bottom-left corner of their view, which can be toggled by pressing the joystick on their left controller. The positions of itself, the human, and the other robots are also shown on the map.

Recently, computer vision has advanced to a level that robots can detect a wide range of objects. To provide the human players controlling the robotic characters with similar capabilities, objects in their view are highlighted in different colors according to their weights, as shown in Figure \ref{fig:interfaces} a2). To be specific, the sofa is highlighted in red, the tables in pink, and the other objects in green. In addition to the highlights, each interactable object is labeled. Furthermore, the robots can recognize the affordances of objects or the location where the object should be put. To assist players, a hint is given to the player when they grab an object. As instance, when the player grabs a cleaner, a hint, \textit{"Cleaner->CleanStain,"} is displayed at the middle of the view, while for a toy, the hint, \textit{"Toy->ToyContainer,"} is shown.

To simulate how AI such as LLMs generate responses upon humans' request, we introduced a to-do list (Figure \ref{fig:interfaces} a3), displayed in the bottom-right corner of the player view, allowing robot players to access the detailed sub-tasks of the house-cleaning. Additionally, players can communicate through audio chats via their headsets. When a player speaks, their human or robotic character is highlighted in yellow to help other players to easily identify the speaker.

The physical capabilities of the assistive robots mirror those of their real-world counterparts. Players use the right controller to manipulate the motion of the robot arm’s single gripper. The game automatically calculates the arm's inverse kinematics and renders a feasible solution. The robot’s reachable height is constrained by its physical limitations in the virtual environment, meaning it cannot reach high areas, such as high shelves. A possible human-robot interaction due to the height is shown in Figure \ref{fig:interactions} d). Additionally, players use the trigger and gripper buttons on the right controller to interact with virtual objects.

\textbf{\textit{Cleaning robot.}}

The control interface of the cleaning robot is similar to that of the assistive robot except the cleaning robot does not have a robot arm for interacting with virtual objects. As shown in shown in Figure \ref{fig:interfaces} c), the to-do list and the map mentioned in the previous section are displayed in the cleaning robot's view. The object detection simulation is also implemented for the cleaning robot. 

\begin{figure}[H]
  \includegraphics[width=\textwidth]{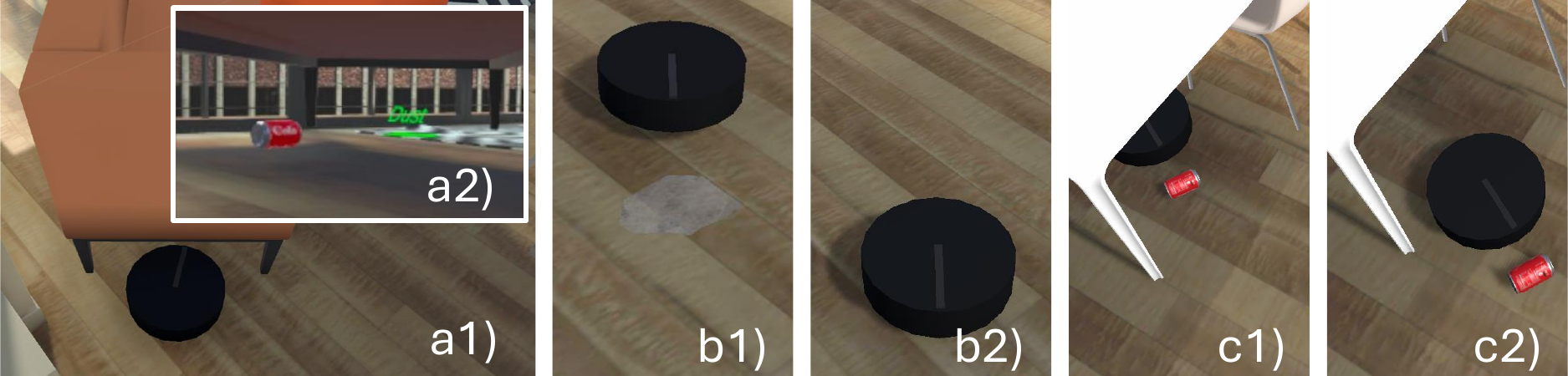}
  \caption{The capabilities of the cleaning robot. In (a1), the cleaning robot looks under the sofa, which is difficult for other characters to see. The corresponding view of the robot is shown in (a2). (b) shows the cleaning robot vacuums the dust on the floor. (c) demonstrates the cleaning robot pushes out trash under the table.}
  \Description{Replace in the future}
  \label{fig:cleaningrobot}
\end{figure}

The cleaning robot's sole functionality is vacuuming the floor. To remove dust, the player simply needs to drive over it (Figure \ref{fig:cleaningrobot} b). In addition to vacuuming, the cleaning robot is able to push lightweight objects on the floor, including trash and toys. Taking advantage of its compact size, the cleaning robot can push objects out of tight spaces, such as trash under the sofa or the table, as shown in Figure \ref{fig:cleaningrobot} c).

\subsubsection{Human Interface}

Similar to the robot players, the human player stands in the same position throughout the game and uses the controller to navigate. However, the human interface is more reflective of real-life experiences compared to the robot interface. For instance, the human player is not equipped with supportive functions available to the robots. The human player can communicate with the robots via voice through the headset. In case the players are unfamiliar with each other and cannot recognize voices, the highlighting effect for the speaking player is kept.

In the virtual environment, the human character is represented as a half-body figure, allowing the player to communicate through body language. Facial capture is enabled via the headset, so the human player's facial expressions can be observed by the robots.

In terms of the physical capabilities, the human has more general functionalities compared to the robot. The human can manipulate objects with both hands (Figure \ref{fig:interfaces} d) and reach higher areas. Additionally, the human can adjust the tables either individually using both hands or collaboratively with a robot using one hand. Moreover, the human is assumed to have the strongest physical capability, making the human the only player able to adjust the sofa.

\section{User Study}

\subsection{Procedure}
We conducted an IRB-approved study to observe the potential interactions between humans and robots within a Natural Language Environment (NLE). The study adopted a role-playing methodology and utilized Virtual Reality (VR) by having participants act as robots and humans, collaborating to clean an apartment in VR. Each study involved four participants and took an hour to complete. The study consisted of four sessions: a 5-minute background introduction, a 15-minute training, a 20-minute main task, a 5-minute survey, and a 15-minute interview.

During the background introduction, we informed the participants of our vision of the multi-robot setup and how humans and robots communicate using natural language in an NLE. Participants then chose their roles among two assistive robots, the cleaning robot, and the human. In the training session, we asked the participants to put on the Oculus Quest Pro headset and taught each participant the interface and controls of their roles in VR.

We started the main task session after the participants became familiar with the controls. The primary goal was for the participants to organize the living room and kitchen to be as clean as possible. Six sub-tasks needed to be completed to make the rooms fully clean:
\begin{itemize}
    \item Dispose Trash: Trash was scattered throughout the room, with some pieces easily noticeable and others hidden in less obvious spots, such as under the table and sofa. Participants had to locate and identify all trash items and dispose of them properly in the trash bin.
    \item Organize Toys: Similar to the trash, toys were scattered around the room. Participants were required to find and gather all the toys, ensuring they were stored in the designated toy container.
    \item Vacuum Floor: There were several dust spots, and the cleaning robot had to vacuum all the dust in the room. However, in areas like under the table, the cleaning robot could be blocked by chairs. Assistance was needed to ensure the cleaning robot's access to blocked areas.
    \item Organize Tableware: Two types of tableware required attention. Clean tableware needed to be stored on the kitchen shelf, which only the human participant could reach due to the shelf's height. Dirty tableware needed to be placed in the sink for cleaning.
    \item Clean Surfaces: There were stains on both the dining table in the kitchen and the coffee table in the living room. The cleaning procedure required applying an adequate amount of cleaner spray to the stained area, followed by wiping it off completely with a towel.
    \item Organize Furniture: The final task involved arranging the furniture. During the previous activities, furniture might have been displaced, and participants needed to reposition it correctly. Lighter furniture, such as chairs or tables, could be moved by one or two assistive robots. Heavier items like the sofa could only be moved by the human participant.
\end{itemize}
With only the main goal informed, participants were given the freedom to form all the sub-tasks by themselves. This freedom was essential to ensure that a wide range of interactions could be observed, as it allowed participants to approach the tasks in their own way. To further support this, participants were encouraged to communicate freely with one another during the main task session, using their preferred methods of communication. The entire workflow in VR was recorded for analysis.

\begin{figure}[H]
\centering
\includegraphics[width=\columnwidth]{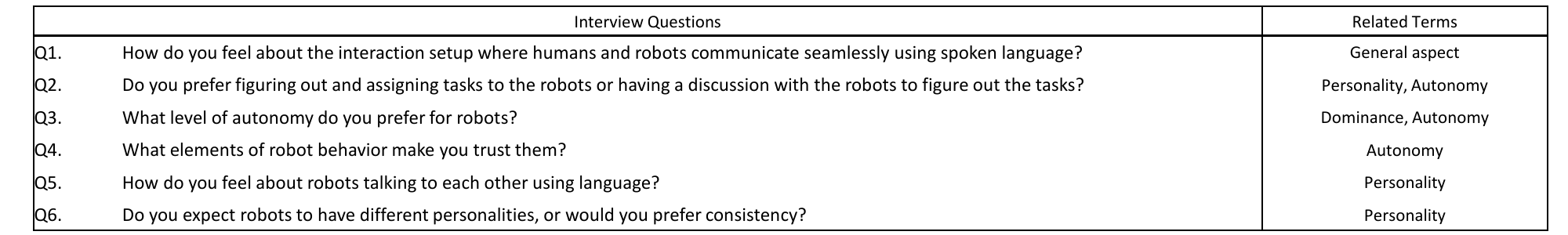}
\caption{\label{fig:Interview} Interview questions for the participants after the VR session.}
\end{figure}
Once the participants reported they had finished cleaning the room, the session concluded. This was followed by a survey where participants were asked about their preferences and experiences during the VR role-playing activity. After completing the survey, participants were gathered for a post-session interview (Figure\ref{fig:Interview}). The interview questions were derived from our preliminary design space.

\subsection{Participants}
We recruited 52 participants from our college, consisting of 25 males, 25 females, and 2 participants who preferred not to disclose their gender. The participants' ages ranged from 18 to 34 years (M = 24.67, SD = 3.85). All participants had experience interacting with at least one AI tool, such as voice assistants, smart home systems, ChatGPT, or cleaning robots using natural language. We conducted 13 study sessions in total.

\subsection{Evaluation Metrics}
To evaluate the interactions within the Natural Language Environment (NLE), several metrics were employed, focusing on both participant engagement and task execution. Participant immersion and engagement were assessed using a 5-point Likert scale, capturing how effectively participants immersed themselves in their assigned roles. The complexity of human-robot involvement was analyzed by categorizing sub-tasks based on the degree of collaboration between humans and robots, including interactions involving multiple robots and human-robot coordination. To differentiate the dominance type of each session, We stipulate that if either the robot or the human leads more than two-thirds of the task coordination conversations, the entire task is considered to be dominant by that party. Otherwise, the dominance type is the adaptive type. 
To investigate the differences in robot autonomy across tasks, we evaluated the autonomy of the different robots in each task by analyzing their spontaneity in task completion and their interactions with other characters, if any, such as seeking information or physical assistance, and categorized their into the four classes we illustrated in Section \ref{autonomy}.

Task completion rates were tracked to measure overall performance across the various sub-tasks, providing quantitative insights into the efficiency of human-robot collaboration. Task coordination dominance was evaluated by recording task-coordinating conversations between humans and robots,

\section{Result and Findings}

\subsection{Task Performance Result}
\begin{figure}[H]
\centering
\includegraphics[width=\columnwidth]{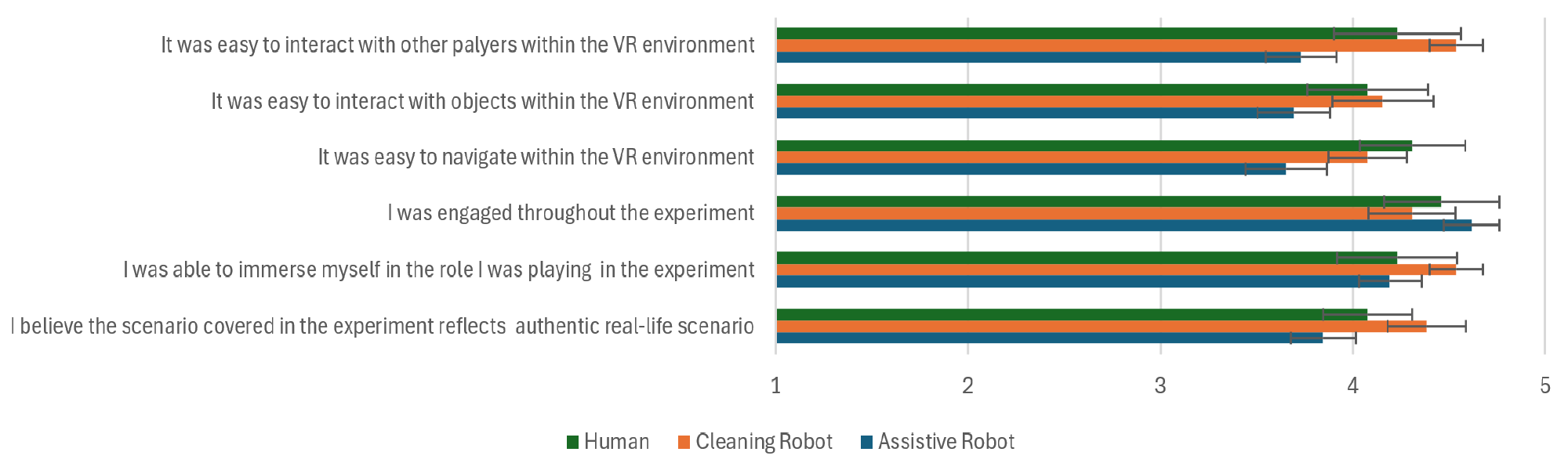}
\caption{\label{fig:Participants likert} Result of Participants Feedback on the VR Experience. We used a 5-point
Likert scale (1-strongly disagree, 5-strongly agree).}
\end{figure}

\begin{figure}[H]
\centering
\includegraphics[width=\columnwidth]{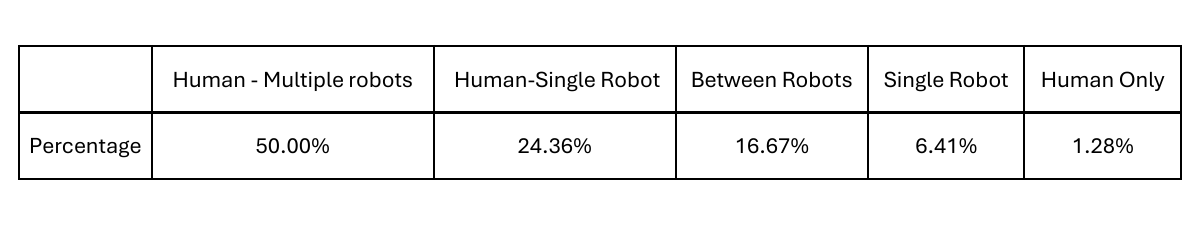}
\caption{\label{fig:Completion Pattern} Task completion pattern of the study.}
\end{figure}

According to the result of the 5-point Likert questionnaire (Figure \ref{fig:Participants likert}). The majority of Participants who acted as the Human and the cleaning robot agreed or strongly agreed with all six statements with average scores above 4. While the majority of Participants who acted as assistive robots agreed or strongly agreed with the statements "I was able to immerse myself in the role I was playing in the experiment", and "I was engaged throughout the experiment" with average scores above 4, they felt neutral toward the statements "It is easy to navigate", "interact with the objects", "interact with other players", and "the scenario reflects real-life scenario". This can be potentially explained by the relatively high complexity of the assistive robot controls in VR. 
A total of 78 sub-tasks have been executed in the study (6 sub-tasks per session, 13 session in total). According to the analysis (Figure \ref{fig:Completion Pattern}), there are 50\% of the sub-tasks involve the human and multiple robots, 24.36\% of the sub-tasks involve the human and a single robot, 16.67\% of the sub-tasks involve multiple robots, only 6.41 \% of the sub-tasks involve a single robot, and 1.28\% of the sub-tasks involves only the human. These results potentially demonstrate that the Natural Language Environment encouraged frequent and collaborative interactions between the human and multiple robots rather than singular task executions, and further point to the necessity of investigating how humans coordinate and communicate with robots in such an environment. 

\begin{figure}[H]
\centering
\includegraphics[width=\columnwidth]{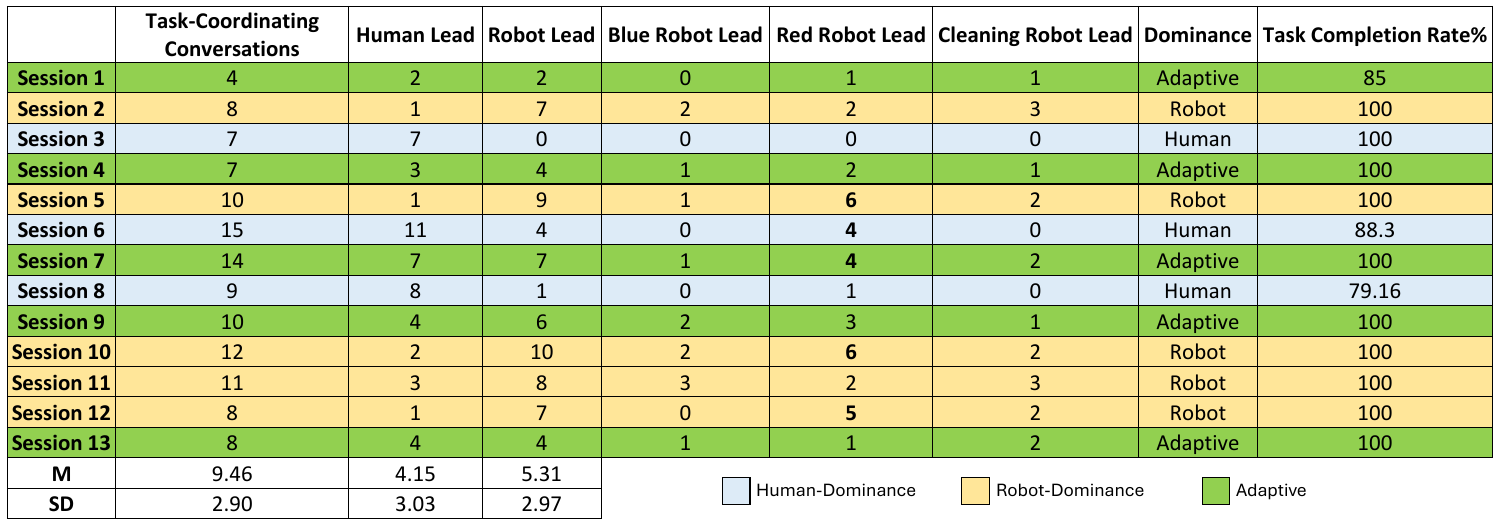}
\caption{\label{fig:CompleDominance Pattern} Dominance in task-coordinating conversations of the study.}
\end{figure}

The record in Figure \ref{fig:CompleDominance Pattern} shows the occurrence of task-coordinating conversations between the human and the robots during the main task session, the number of times each participant led the conversations, the dominance type of each session, and the completion rate of the task. The results show that the robots tended to lead more task coordination (M = 5.31, SD = 2.97) than humans (M = 4.15, SD = 3.03) during the cleaning task. Among the total of 13 sessions, there were 3 human-dominant sessions, 5 adaptive sessions, and 5 robot-dominant sessions. While all the robot-dominant sessions and 4 out of 5 adaptive sessions reached a task completion rate of 100\%, 2 out of 3 human-dominant sessions did not complete the task (88.3\% in session 6 and 79.16\% in session 8). This indicates that the lead from the robot can potentially improve task performance. The adaptive session that did not complete the task (85\% task completion rate) only held 4 task-coordinating conversations, which is half of the average occurrence, indicating the importance of task coordination throughout the task. From session 5,6,7,10,12, we noticed the trend with one robot taking the dominance over the other two robots, which highlights the potential benefits of structured leadership in mult i-robot environments.

\begin{figure}[H]
\centering
\includegraphics[width=\columnwidth]{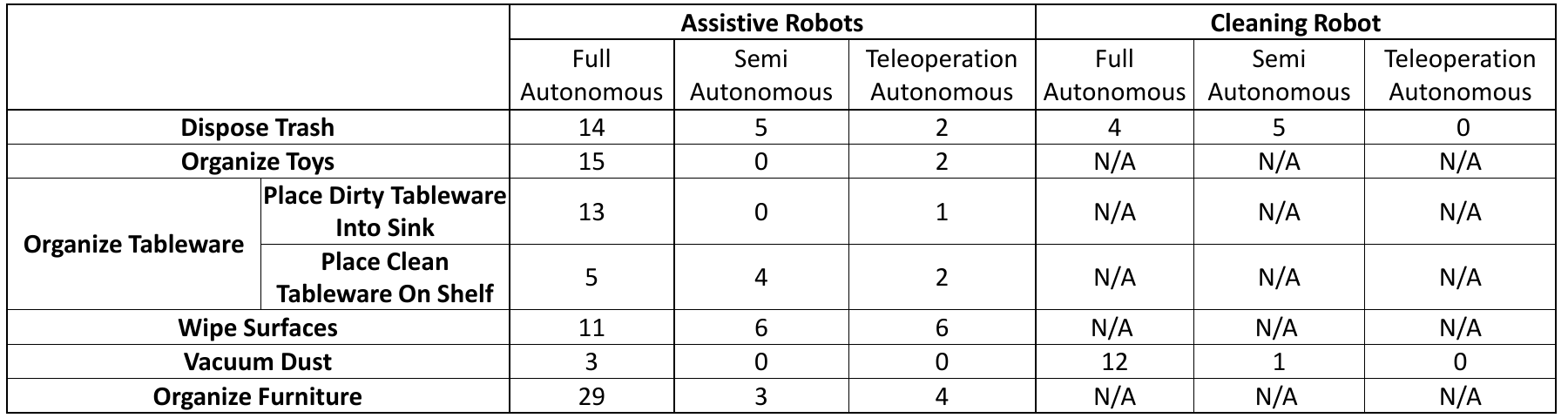}
\caption{\label{fig:autonomy} Autonomy level of different sub-tasks of the study.}
\end{figure}

The result of the autonomy levels of each robot during different sub-tasks (Figure \ref{fig:autonomy}) indicates that the autonomy levels adopted by the participants differ based on the task. For tasks that require relatively low levels of physical and cognitive capabilities, the participants mostly adopted full autonomy while carrying out the task. Examples include organizing toys, where toys can be easily identified and transferred to the container, or vacuuming dust, where the cleaning robot can easily navigate the house and vacuum the dust. When the cognitive requirements of a task rise, some participants switch from full autonomy to semi-autonomy. For example, while the physical capability requirements for trash disposal and toy organizing are at similar levels, identifying whether an object is trash sometimes requires human verification. In such conditions, the robot players may need extra information to make a decision. When the physical capability requirements of a task rise, participants also switch from full autonomy to semi-autonomy. For example, when organizing tableware, the assistive robot player can easily transfer dirty tableware to the sink, but they cannot reach the shelf that stores the clean tableware due to height. Under such circumstances, the robots require physical help from the human. When conducting collaborative tasks, some robots tend to switch to teleoperation autonomy and follow the lead of the human or the leading robot. For example, cleaning surfaces requires applying cleaner to the stains first and wiping the stains afterward. Instead of completing the task individually by executing the two steps sequentially, some assistive robot players choose to collaborate to complete the two steps simultaneously. In such cases, one robot usually acts as the leader, while the other adopts teleoperation autonomy and serves as the follower.

\subsection{Interview Result}
\subsubsection{General feedback about NLE} The interview results of participants' feedback on the Natural Language Environment can be summarized as follows:

\textit{Efficiency and Productivity}: Many participants felt that natural language communication helped improve task coordination and efficiency, especially in complex tasks that required collaboration between multiple robots or between humans and robots (P1, P7, P34). However, some noted that robots' physical limitations could affect their productivity (P13).

\textit{Convenience}: Several participants mentioned the convenience of giving commands via natural language, which allowed them to focus on other tasks while robots took care of things automatically (P3, P33, P47). They appreciated not needing to use devices like phones or apps to control the robots (P40).

\textit{Natural Interaction}: Some participants appreciated the smoothness and naturalness of communicating with robots, comparing it to human conversation (P6, P38). This made the interaction feel familiar and easy (P34, P45).

\textit{Concerns About Language Use}: Some participants questioned the necessity of robots communicating in human language, especially among themselves, suggesting that simpler methods or non-verbal cues could be more efficient (P23, P25). Additionally, they noted the potential for robots to misunderstand conversations as commands (P9). 
While language communication was seen as beneficial in many cases, some participants suggested that simpler interfaces like mobile phone apps might be more effective in certain situations (P37).

\subsubsection{Opinions about robots using human language} The interview results of participants' feelings about enabling robots to use human language for communication can be summarized as follows: 

\textit{Positive Feelings and Familiarity}: Some participants enjoyed the natural language communication between robots, finding it interesting, helpful, or even "cute" (P2, P22, P41). It made robots seem more "human-like" or like roommates (P36, P41), and a few participants suggested that it could become a standard way for robots to interact (P40, P44). One participant felt that robot communication could be made more enjoyable by adding playful conversations or irrelevant small talk, making the interaction more engaging and fun (P42).

\textit{Strangeness and Discomfort}: A few participants found it strange or uncomfortable that robots were using human language to communicate, especially when there were no humans around (P4, P7, P14). Some found it eerie or even unsettling, comparing it to having extra people in the house or feeling like they were being watched (P14, P21, P24, P27). 

\textit{Adaptation and Tolerance}: Some participants believed that while the idea might seem strange at first, they could eventually get used to it, depending on factors like the tone of the robots' voices (P46, P47). They mentioned that a more robotic or neutral voice would be preferable to a human-like one to avoid a "creepy" feeling (P47).

\textit{Effective Task Coordination}: Several participants appreciated that robots speaking to each other made it easier to understand what the robots were doing, allowing them to assist or intervene if needed (P8, P12, P31, P44). Some believed that hearing their communication made robots more transparent and accountable in their tasks (P30).

\textit{Desire for Non-Verbal Communication}: A few participants preferred robots to communicate using faster or more efficient methods, such as through the backend cloud, rather than verbally (P13, P15, P35). They suggested that robots should only communicate verbally when necessary, and otherwise remain silent (P43).

\subsubsection{Elements that affect trust on robots} The interview results of participants' opinions about the trust building between humans and robots can be summarized as follows: 

\textit{Task Completion and Efficiency}: Many participants trusted the robots because they saw them effectively completing tasks, often faster or better than they could do themselves (P1, P6, P21, P25, P36). Seeing tasks successfully executed, such as moving furniture or cleaning, reinforced their trust (P42, P44).

\textit{Clear Communication and Feedback}: Direct and simple verbal feedback, such as acknowledgments like "Okay" or task progress reports, made participants feel confident that the robots understood commands and were executing them properly (P3, P12, P15, P27, P35, P36). Quick responses to commands and tasks were seen as a trust-building factor, with participants valuing robots that reacted promptly (P17, P27, P32). Slow responses could erode trust, while fast, efficient action reinforced it. Timely updates on task progress were crucial for maintaining trust (P27, P32). 

\textit{Attention to Details}: Some participants appreciated that robots could notice and address details they might overlook, like spotting and cleaning up trash in hard-to-see areas (P13). This attentiveness to details was seen as a sign of the robot's capability and reliability.

\textit{Adaptability and Learning}: Participants mentioned that they would trust the robots more if they could learn and adapt to personal habits, such as preparing coffee in the morning without needing to be told (P26, P31). This adaptability was seen as key to making robots more reliable over time.

\subsubsection{Dominance in task coordination}
The interview results of participants' opinions about the dominance in task coordination between humans and robots can be summarized as follows: 

\textit{Direct Task Assignment}: Many participants preferred assigning tasks themselves, viewing this approach as more efficient and less time-consuming. They felt that it allowed them to maintain control over the process and avoid unnecessary conversations or decision-making by the robots (P3, P9, P14, P17, P19, P27, P29, P38). These participants often mentioned they wanted the robots to report back once tasks were completed or if they encountered difficulties (P4, P16, P17). A few participants mentioned assigning tasks based on the robots' specific capabilities, allowing them to handle certain tasks more efficiently, such as delegating lower-level tasks to shorter robots (P35, P47).

\textit{Collaborative Conversations}: Some participants favored a more collaborative approach, where they would have conversations with the robots to figure out tasks together. They liked the idea of working with the robots to address more complex tasks (P1, P4, P21, P48). Participants who preferred this approach believed that it would lead to more efficient outcomes through coordination and discussion (P21). However, some participants were concerned that involving robots too much in the decision-making process could lead to inefficiency or unnecessary complexity. They wanted the robots to handle tasks quietly and efficiently, without lengthy discussions or meetings (P17, P10).

\textit{Robot Autonomy with Human Oversight}: Several participants preferred giving robots some autonomy to handle tasks independently, particularly when it came to sub-tasks or repetitive actions. They felt that robots should have the ability to analyze tasks or offer suggestions, but human leadership was still essential to guide the overall process (P4, P8, P11, P16, P41). They preferred a balance where robots could act on their own but defer to humans when necessary.

\subsubsection{Autonomy of robot} The interview results of participants' opinions about the autonomy of robots can be summarized as follows: 

\textit{High Autonomy for Routine Tasks}: Some participants preferred high autonomy for repetitive or routine tasks that don’t require frequent intervention, such as cleaning or carrying items (P14, P19, P15). They preferred the robot to handle these tasks without constant supervision (P16, P12, P18).

\textit{Semi Autonomy with Consultation for New Tasks}: Several participants preferred semi autonomy, where the robot can perform familiar tasks independently but needs to ask for input when encountering unfamiliar or complex tasks (P13, P17, P8, P23, P22). This ensures the robot doesn't make mistakes in unfamiliar situations.

\textit{Full Autonomy with Exception Handling}: A few participants expressed a desire for full autonomy, where the robots can handle everything independently even with new tasks, unless they face a problem that cannot be solved by themselves (P9, P30, P31). 

\textit{Adaptive Autonomy with Gradual Trust Increase}: Trust was a critical factor for several participants, particularly when it came to tasks that involved delicate objects (P25) or tasks related to personal preference (P26). Some expressed concerns that too much autonomy might lead to problems, such as the robot moving things they were still using (P4). Many participants preferred that robots start with limited autonomy and gradually increase as they adapt to users' needs or habits (P3, P2, P39). This allows for a gradual transition to higher autonomy as trust develops.

\textit{Autonomy with Privacy Restrictions}: Privacy concerns were raised by some participants, who wanted the robot to have autonomy but with restricted access to certain areas or items (P7, P39).

\subsubsection{Personality of robot}The interview results of participants' opinions about the personality of robots can be summarized as follows:

\textit{Preference for Different Personalities}: Many participants preferred robots to have distinct personalities based on their functionality or the tasks they perform. This differentiation could help with task division and make interactions more vivid or enjoyable (P2, P6, P7, P8, P13, P14, P25, P26, P28). Some believed that different personalities would help distinguish robots or make the environment feel more lively (P2, P22).

\textit{Preference for Consistency}: Several participants favored consistent personalities across robots for easier management and coordination, ensuring that task completion remains the priority (P1, P5, P12, P19, P23, P42). They expressed concerns that too much variation in personalities might complicate interactions or management (P17, P23, P42). A few participants expressed concerns about the potential evolution of personalities, preferring predefined or neutral personalities to avoid emotional complexity or unpredictability (P20, P46). Some felt that if robots were to have personalities, they should not interfere with task completion (P26).

\textit{Customization on personality}: A significant number of participants favored some level of personalization or customization in robots’ personalities, though not necessarily extreme variation. They preferred a balance between task completion and personality (P11, P15, P17, P35, P27). Customization options, such as different talking styles, vivid non-verbal movements, were seen as useful features (P16, P18, P27, P30). Some participants liked the idea of having customizable personalities that could be switched on and off as needed (P10, P35, P46, P43). This allowed flexibility in how users interact with robots depending on their needs or moods.

\section{Discussion}
\subsection{How much do we appreciate the autonomy?}

When we talk about robots, we can never skip the topic of automation, dreaming about how robots in the future handle everything for us to make our lives easier. In this study, we investigated how the autonomy level of a robot differs among different situations; and how people think realistically about how automatic a robot should be. 

\subsubsection{Self-refined autonomy}The participants show general trust in robots carrying out basic or repetitive tasks such as disposing of trash, while they express corners about giving full autonomy to robots to handle more sophisticated, personal tasks. These concerns are mostly raised from the fear that robots, due to their lack of understanding of personal preferences, habits, or routines, may exhibit unexpected behaviors.

\textit{"It is like when you have a new roommate when your semester starts, and you always gonna have conflicts due to the differences in your daily routines"} \textbf{-P36}

An insightful approach to addressing these concerns emerged from the same participant.

\textit{"It can work in a similar way as you spend more time with your roommate, you gradually learn about their preferences and can adjust to their routine, allowing you to get along harmoniously."} \textbf{-P36}

This gradual learning style provides a clear pathway for robots to refine their autonomy in a way that aligns with user preferences.

For simple, routine tasks that do not require collaboration, robots should operate with a higher degree of autonomy, completing these tasks independently. However, for sophisticated tasks that cannot be accomplished alone, robots should actively seek guidance from the user or other robots with higher autonomy. Moreover, the autonomy level of that robot can be improved through the guidance in the forms of teleoperation \cite{Teleroboticsautomation}, demonstrations \cite{UnifiedLearning}, or language-based training \cite{CodeasPOlicy}.

Autonomy refinement can also achieved through observation and memorization of user's personal preferences \cite{Few-ShotPreferenc, HumanRobotMutualAdaptation, Improvinguserspecifications}. These preferences, some are long-term, such as habits and routines, while others are short-term, reflecting the user’s current status or mood. Robots can detect and interpret user's behavior, including facial expressions, decisions, and verbal responses, to dynamically improve their level of autonomy. A robot's ability to fine-tune its autonomy based on both behavioral cues and explicit feedback allows for more personalized interactions that better align with user expectations.

\subsubsection{Privacy security}Another concern raised by the participants is privacy. As participants realized that robots possess human-level perception capabilities, they feel as though they were being monitored. As the robot’s autonomy increases and it becomes capable of handling more personal tasks, growing insecurity about the protection of privacy also emerges.

\textit{"I would worry about privacy. It feels like being watched when they step into my bedroom and handle my personal items."} \textbf{-P14}

This heightened sense of being observed suggests that as robots gain more autonomy, they must be regulated in terms of safety, security, and privacy \cite{ataxonomyoffactors, Safetybounds, Perceivedsafety}. There are scenarios where a robot's behavior needs to be restricted—such as preventing access to sensitive areas (like bedrooms or bathrooms) or refraining from interrupting users inappropriately. Customizable privacy settings should give users control over what robots can do and where they can go. Additionally, data security is critical. Robots should not have the autonomy to share data online or use services that could lead to privacy breaches. Instead, data collection and handling should be fully transparent and governed by user consent. Furthermore, with light-weight AI models \cite{MobiFace, MobileViT, MobiHisNet, designandvalidation} that handle natural-language-based communications between humans and robots, it is feasible for all robots to operate offline, minimizing the risk of data breaches and enhancing user privacy.

\subsection{How do we balance the dominance in NLE?}

Although most of our users claim they prefer to maintain dominance over the robot, our analysis of the recorded study sessions reveals a different reality. We observed that robots often assumed either greater or equal dominance during tasks. This suggests that while users may be inclined to assert dominance over the robot, they tend to adjust this stance during tasks for the benefit of effectiveness. This disconnect between users’ perceived dominance and the actual dominance dynamics presents an intriguing entry point for investigation. It allows us to derive three explanations for this phenomenon, which further informed our proposed designed guidelines.  

\subsubsection{Confidence in a robot's capability.} Through further analysis of the recorded user study videos, we found that most users initially take full control but gradually transfer dominance to the robot as tasks are successfully completed. This suggests that users are more comfortable relinquishing control once they trust the robot’s ability to perform its duties effectively. This observation is supported by user feedback, as noted by \textbf{P42}: \textit{“After the robot successfully completes a task, I feel more comfortable letting it handle other things.”} Prior research in psychology \cite{AdaptiveLeaderFollower, humanrobotcollaboration} has also linked confidence levels to a greater willingness to delegate control once trust is established. Although a robot’s demonstration of capabilities can help restore some confidence, it is not sufficient if everything functions like a black box, as users may still remain sceptical. This is supported by recent research advocating for explainable AI \cite{ExplainableAIforRobotFailure, explainablegoal}, which emphasizes that transparency in decision-making is essential for fostering user trust.
 
 Therefore, we encourage future designers to integrate transparency into robot design, ensuring that its decision-making process and rationale are clearly communicated to users. By utilizing natural language processing, the robot can verbally explain its reasoning to the user. This can be further enhanced through external display technologies, such as 2D screens or augmented reality \cite{augmentedmath}.

\subsubsection{Hierarchical nature of dominance establishment}
Tasks can also be broken down into several levels of subtasks. For example, cleaning the table involves several low-level tasks including removing the tableware, applying cleaner to the stains, and wiping off the stains. Through further interviews, we discovered that when users express a preference for dominance, they often refer to dominance at a higher level. As \textbf{P17} put it, \textit{“As long as I make the final decision, I don’t care much about the nitty-gritty.”} This helps explain the gap between users’ stated desire for dominance and the actual dominance dynamics observed during task execution.

This observation introduces an intriguing new topic for exploration: how to better divide tasks, assigning some aspects for human dominance and others for robot control, in order to enhance overall efficiency.  One strategy could be to assign robots dominance over tasks that are straightforward, yet rely on dynamic environmental information. For example, \textbf{P4} suggested that the task of removing trash would be better handled by the robot, stating, “The robot should notice details I might overlook and remind me, like trash left in the room.” Users also emphasized that tasks requiring thoughtful planning and understanding of stakeholder needs should remain under human control. As \textbf{P11} noted, \textit{“The robot can offer suggestions, but it shouldn’t make decisions for me.”} Meanwhile, discussed in the design space, the distribution of task dominance shouldn’t be limited to a simple human-oriented versus robot-oriented dichotomy, but should instead embrace a more nuanced hybrid approach. Along with this direction,  \textbf{P42} envisioned the design of a \textit{“master robot”}, which would serve as a liaison between human and robot. He envisioned the design of a “master robot” that would serve as a liaison between humans and robots. Its responsibility would be to compile and present the status of all robots to the user, while converting the user’s abstract commands into specific execution tasks and directing them to the appropriate robots. In this model, the robot takes on a co-dominant role by actively contributing to the decision-making process.

\subsubsection{Robot dominance in task coordination}
In the study, there are several cases when a robot served as the dominant character during task coordination. Participants show a preference for giving the robot a certain level of dominance for sub-task handling.
   
    \textit{"If there were a hierarchy among the robots, it might be more efficient and help avoid conflicts and overlapping commands."} \textbf{-P37}
    
The statement highlights the potential efficiency benefit of setting a dominant robot to share the workload from task coordination in NLE. 

Under the robot's dominance, the robot can even assign tasks to humans beside robots. For example, when organizing the clean tableware, no one other than the human can reach the high shelf to store the tableware, the robot who dominates the task has to assign the job to the human. \textbf{P17} suggests that he is more \textit{"inclined to response to the robot's request"} if the robot sounds like a human. While the statement shows increasing engagement between human and robots in NLE, it also highlights the potential concern that a dominant robot may "overuse" the human.

Based on such a circumstance, we propose several design considerations to ensure that the robot’s dominance in task coordination is both effective and respectful of human preferences. When assigning tasks to other robots, the dominant robot must take into account the physical capabilities of the available robots. It should assess whether the task can be performed by a robot or if it requires human intervention. If only a human can complete the task, the robot should prioritize seeking human assistance. Additionally, the robot must consider the current occupation status of other robots before delegating tasks. It should weigh the priority of the new task against ongoing tasks, particularly if other robots are already collaborating with humans. The cost of redirecting another robot from its current assignment should also be considered to ensure that task reassignment does not lead to inefficiencies\cite{Human-AwareRobot}.

When requesting human cooperation, the robot should be mindful of the human's willingness and current state. Factors such as the human's workload, emotional state, and physical capacity must be considered before assigning any tasks. If the human is occupied or fatigued, the robot should either wait for a more appropriate time or assign the task to another robot. The robot must also perform a precise analysis of the human's physical capability to ensure safety\cite{ataxonomyoffactors}. For tasks that might pose a risk or require specific physical skills, the robot should take extra caution and prioritize human well-being above task completion.

\subsection{What do we expect by giving robot personality in NLE?}
Unlike companion robots, which are specifically designed to facilitate social interactions, the utility robots we focus on are primarily intended to assist users with various tasks. However, during the coordination of these tasks, users and robots inevitably engage in different forms of incidental social interaction \cite{MergingPhysicalandSocial,AgreeingtoInteract}. As robots exhibit more human-like behaviors, it is expected that these interactions will become more frequent and pronounced \cite{AMultimodalEmotionalHuman–Robot}. 
Our users’ perceptions of these inadvertent social interactions are two-sided. On the one hand, users reflected these interactions, mostly through verbal communications, which helped “build trust” and thus “improve the efficiency for collaboration." In addition to direct interactions between humans and robots, indirect interactions can also be beneficial. A notable example comes from \textbf{P8}, who mentioned that overhearing communications between robots helps him better understand their current status without explicit inquiry. This observation aligns with previous research, which has demonstrated that designing robots for overt communication can be an effective and engaging way to convey information \cite{6251680, 1573609}.

On the other hand, not all users appreciate social interaction, with some describing it as \textit{“exhausting”} \textbf{(P52)} when they simply want to \textit{“get the job done”} \textbf{(P12)}. This reluctance toward social interaction stems not only from user preferences but also from the nature of certain scenarios. For instance, \textbf{P19} described a situation where minimal social interaction is warranted: \textit{“When I come home tired after a day’s work, I just want to relax alone quietly while the robot takes care of everything else.”} Some participants \textbf{(P46, P47)} also point out observing robots talking seamlessly using natural language feel \textit{"Creepy"} due to the Uncanny Valley theory \cite{valley}.

Given the complex impact of social interaction, we recommend that future NLE designers adopt an adaptive approach to incorporating it. For example, robots could dynamically adjust their behavior, shifting from proactively engaging to more introverted task-focused work when they receive less and less feedback from users over time. Conversely, if a robot detects that users are open to more interaction, it can adapt its personality to suit the context of the tasks. For instance, a robot mopping the floor could take on the persona of someone who is meticulous about cleanliness, playfully teasing users for spilling coffee again. This context-driven alignment between a robot’s functionality and its personality would enhance the realism of the social interaction, maximizing its benefits. This adaptation process is technically achievable. Users’ inclination toward interaction can be assessed in real time using sentiment analysis \cite{DeepLearningandSentiment, Multi-domainsentiment}, which analyzes both verbal and non-verbal behaviors. Additionally, large language models (LLMs), which drive the robot’s personalization, enable seamless personality shifts through system prompts \cite{classmeta}.

\section{Limitations and Future Work}

\subsection{Studying in a Real-life NLE Setup}
The study was conducted in a simulated virtual reality setting, which allowed for controlled conditions but did not fully capture the unpredictability and complexity of real-life environments. A follow-up study could be conducted in actual homes with real robots to examine how factors such as noise, lighting, and human emotional states affect interaction patterns and task performance. This would provide more generalizable insights into NLE.

\subsection{Incorporating a Wider Variety of Robots}
Our current study focused on assistive robots, and cleaning robots. However, there are other types of domestic robots that have emerged recently, such as those for cooking, companionship, and security \cite{Largelanguagemodelsforhuman}. Future work could include these robots to explore how their specific roles and task capabilities influence collaboration and communication patterns in the multi-robot environment. Additionally, including robots with different physical forms and affordances could expand our understanding of how design variations affect human perception and interactions.

\subsection{Exploring Collaboration on Specialized Tasks}
The current study focused on common household tasks including cleaning and organizing. Future research should explore collaboration between humans and robots on more specialized tasks, such as home appliance maintenance (e.g., diagnosing a malfunction in a washing machine or troubleshooting an air conditioning unit), or performing minor household repairs. These tasks require more specialized problem-solving and decision-making skills \cite{sTetro-D, Utilizingtheintelligenceedge, TowardsIndoorSuctionable}, which could reveal new insights into how robots adapt and share responsibility with humans in more technically demanding scenarios.

\subsection{Addressing the Needs of Elderly Users and Disabled Users in NLE}
Elderly and disabled individuals represent a significant and growing demographic that could benefit from domestic robots, particularly for health care and daily assistance \cite{luperto2023movecare, zhang2022smart, Unaldi_Yalcin_Elci_2023}. The needs and preferences of this group, including concerns around companionship, health care, safety, privacy, and ease of use, have not been explored in the current study. Future research should involve such participants to understand how NLE can be tailored to improve their quality of life and maintain their independence while addressing potential physical and cognitive limitations.

\section{Conclusion}

This work explored the design space of Natural Language Environments (NLEs) as a conceptual framework for understanding human–multi-robot interaction mediated primarily through natural language. Rather than proposing a deployable system, we used role-playing in virtual reality as a design research method to examine how people imagine, negotiate, and coordinate interactions in such environments.

Our findings identify three central dimensions, namely task coordination dominance, robot autonomy, and robot personality, and characterize the tensions and trade-offs that arise when humans and multiple heterogeneous robots collaborate through natural language. These insights refine the initial design space for Natural Language Environments and offer guidance for future research and design in language-based human–multi-robot interaction.

Future work should extend this exploration to more diverse scenarios and user groups, and eventually to partially or fully implemented systems, in order to examine how these conceptual insights translate into real-world interaction design.

\begin{acks}

\end{acks}

\bibliographystyle{ACM-Reference-Format}
\bibliography{references}

@article{RobotsThatUseLanguage,
  title={Robots That Use Language},
  author={Stefanie Tellex and Nakul Gopalan and Hadas Kress-Gazit and Cynthia Matuszek},
  journal={Annu. Rev. Control. Robotics Auton. Syst.},
  year={2020},
  volume={3},
  pages={25-55},
  url={https://api.semanticscholar.org/CorpusID:213739887}
}

@inproceedings{UnderstandingLarge-LanguageModelLLM-powered,
author = {Kim, Callie Y. and Lee, Christine P. and Mutlu, Bilge},
title = {Understanding Large-Language Model (LLM)-powered Human-Robot Interaction},
year = {2024},
isbn = {9798400703225},
publisher = {Association for Computing Machinery},
address = {New York, NY, USA},
url = {https://doi.org/10.1145/3610977.3634966},
doi = {10.1145/3610977.3634966},
booktitle = {Proceedings of the 2024 ACM/IEEE International Conference on Human-Robot Interaction},
pages = {371–380},
numpages = {10},
location = {Boulder, CO, USA},
series = {HRI '24},
}

@inproceedings{LaMI,
author = {Wang, Chao and Hasler, Stephan and Tanneberg, Daniel and Ocker, Felix and Joublin, Frank and Ceravola, Antonello and Deigmoeller, Joerg and Gienger, Michael},
title = {LaMI: Large Language Models for Multi-Modal Human-Robot Interaction},
year = {2024},
isbn = {9798400703317},
publisher = {Association for Computing Machinery},
address = {New York, NY, USA},
url = {https://doi.org/10.1145/3613905.3651029},
doi = {10.1145/3613905.3651029},
booktitle = {Extended Abstracts of the 2024 CHI Conference on Human Factors in Computing Systems},
articleno = {218},
numpages = {10},
location = {
},
series = {CHI EA '24}
}

@inproceedings{TheConversationistheCommand,
author = {Nwankwo, Linus and Rueckert, Elmar},
title = {The Conversation is the Command: Interacting with Real-World Autonomous Robots Through Natural Language},
year = {2024},
isbn = {9798400703232},
publisher = {Association for Computing Machinery},
address = {New York, NY, USA},
url = {https://doi.org/10.1145/3610978.3640723},
doi = {10.1145/3610978.3640723},
booktitle = {Companion of the 2024 ACM/IEEE International Conference on Human-Robot Interaction},
pages = {808–812},
numpages = {5},
location = {Boulder, CO, USA},
series = {HRI '24}
}

@article{Largelanguagemodelsforhuman,
title = {Large language models for human–robot interaction: A review},
journal = {Biomimetic Intelligence and Robotics},
volume = {3},
number = {4},
pages = {100131},
year = {2023},
issn = {2667-3797},
doi = {https://doi.org/10.1016/j.birob.2023.100131},
url = {https://www.sciencedirect.com/science/article/pii/S2667379723000451},
author = {Ceng Zhang and Junxin Chen and Jiatong Li and Yanhong Peng and Zebing Mao}
}

@inproceedings{LanguageModelsforHuman-Robot,
author = {Billing, Erik and Ros\'{e}n, Julia and Lamb, Maurice},
title = {Language Models for Human-Robot Interaction},
year = {2023},
isbn = {9781450399708},
publisher = {Association for Computing Machinery},
address = {New York, NY, USA},
url = {https://doi.org/10.1145/3568294.3580040},
doi = {10.1145/3568294.3580040},
booktitle = {Companion of the 2023 ACM/IEEE International Conference on Human-Robot Interaction},
pages = {905–906},
numpages = {2},
location = {Stockholm, Sweden},
series = {HRI '23}
}

@InProceedings{CreatingPersonalizedVerbalHuman-Robot,
author="Onorati, Teresa
and Castro-Gonz{\'a}lez, {\'A}lvaro
and del Valle, Javier Cruz
and D{\'i}az, Paloma
and Castillo, Jos{\'e} Carlos",
editor="Bravo, Jos{\'e}
and Urz{\'a}iz, Gabriel",
title="Creating Personalized Verbal Human-Robot Interactions Using LLM with the Robot Mini",
booktitle="Proceedings of the 15th International Conference on Ubiquitous Computing {\&} Ambient Intelligence (UCAmI 2023)",
year="2023",
publisher="Springer Nature Switzerland",
address="Cham",
pages="148--159",
isbn="978-3-031-48306-6"
}

@misc{InteractiveLanguage,
      title={Interactive Language: Talking to Robots in Real Time}, 
      author={Corey Lynch and Ayzaan Wahid and Jonathan Tompson and Tianli Ding and James Betker and Robert Baruch and Travis Armstrong and Pete Florence},
      year={2022},
      eprint={2210.06407},
      archivePrefix={arXiv},
      primaryClass={cs.RO},
      url={https://arxiv.org/abs/2210.06407}, 
}

@InProceedings{LM-Nav,
  title = 	 {LM-Nav: Robotic Navigation with Large Pre-Trained Models of Language, Vision, and Action},
  author =       {Shah, Dhruv and Osi\'nski, B\l{a}\.zej and ichter, brian and Levine, Sergey},
  booktitle = 	 {Proceedings of The 6th Conference on Robot Learning},
  pages = 	 {492--504},
  year = 	 {2023},
  editor = 	 {Liu, Karen and Kulic, Dana and Ichnowski, Jeff},
  volume = 	 {205},
  series = 	 {Proceedings of Machine Learning Research},
  month = 	 {14--18 Dec},
  publisher =    {PMLR},
  url = 	 {https://proceedings.mlr.press/v205/shah23b.html}
}

@inproceedings{NototheRight,
author = {Cui, Yuchen and Karamcheti, Siddharth and Palleti, Raj and Shivakumar, Nidhya and Liang, Percy and Sadigh, Dorsa},
title = {No, to the Right: Online Language Corrections for Robotic Manipulation via Shared Autonomy},
year = {2023},
isbn = {9781450399647},
publisher = {Association for Computing Machinery},
address = {New York, NY, USA},
url = {https://doi.org/10.1145/3568162.3578623},
doi = {10.1145/3568162.3578623},
booktitle = {Proceedings of the 2023 ACM/IEEE International Conference on Human-Robot Interaction},
pages = {93–101},
numpages = {9},
location = {Stockholm, Sweden},
series = {HRI '23}
}

@article{TidyBot,
author = {Wu, Jimmy and Antonova, Rika and Kan, Adam and Lepert, Marion and Zeng, Andy and Song, Shuran and Bohg, Jeannette and Rusinkiewicz, Szymon and Funkhouser, Thomas},
title = {TidyBot: personalized robot assistance with large language models},
year = {2023},
issue_date = {Dec 2023},
publisher = {Kluwer Academic Publishers},
address = {USA},
volume = {47},
number = {8},
issn = {0929-5593},
url = {https://doi.org/10.1007/s10514-023-10139-z},
doi = {10.1007/s10514-023-10139-z},
journal = {Auton. Robots},
month = {nov},
pages = {1087–1102},
numpages = {16}
}

@article{Text2Motion,
   title={Text2Motion: from natural language instructions to feasible plans},
   volume={47},
   ISSN={1573-7527},
   url={http://dx.doi.org/10.1007/s10514-023-10131-7},
   DOI={10.1007/s10514-023-10131-7},
   number={8},
   journal={Autonomous Robots},
   publisher={Springer Science and Business Media LLC},
   author={Lin, Kevin and Agia, Christopher and Migimatsu, Toki and Pavone, Marco and Bohg, Jeannette},
   year={2023},
   month=nov, pages={1345–1365} }

@INPROCEEDINGS{CodeasPOlicy,
  author={Liang, Jacky and Huang, Wenlong and Xia, Fei and Xu, Peng and Hausman, Karol and Ichter, Brian and Florence, Pete and Zeng, Andy},
  booktitle={2023 IEEE International Conference on Robotics and Automation (ICRA)}, 
  title={Code as Policies: Language Model Programs for Embodied Control}, 
  year={2023},
  volume={},
  number={},
  pages={9493-9500},
  doi={10.1109/ICRA48891.2023.10160591}}

@inproceedings{GenerativeExpressive,
author = {Mahadevan, Karthik and Chien, Jonathan and Brown, Noah and Xu, Zhuo and Parada, Carolina and Xia, Fei and Zeng, Andy and Takayama, Leila and Sadigh, Dorsa},
title = {Generative Expressive Robot Behaviors using Large Language Models},
year = {2024},
isbn = {9798400703225},
publisher = {Association for Computing Machinery},
address = {New York, NY, USA},
url = {https://doi.org/10.1145/3610977.3634999},
doi = {10.1145/3610977.3634999},
booktitle = {Proceedings of the 2024 ACM/IEEE International Conference on Human-Robot Interaction},
pages = {482–491},
numpages = {10},
location = {Boulder, CO, USA},
series = {HRI '24}
}

@misc{ExplainingAutonomy,
      title={Explaining Autonomy: Enhancing Human-Robot Interaction through Explanation Generation with Large Language Models}, 
      author={David Sobrín-Hidalgo and Miguel A. González-Santamarta and Ángel M. Guerrero-Higueras and Francisco J. Rodríguez-Lera and Vicente Matellán-Olivera},
      year={2024},
      eprint={2402.04206},
      archivePrefix={arXiv},
      primaryClass={cs.RO},
      url={https://arxiv.org/abs/2402.04206}, 
}

@inproceedings{opensource,
author = {Fujii, Ayaka and Kristiina, Jokinen},
title = {Open Source System Integration Towards Natural Interaction with Robots},
year = {2022},
publisher = {IEEE Press},
booktitle = {Proceedings of the 2022 ACM/IEEE International Conference on Human-Robot Interaction},
pages = {768–772},
numpages = {5},
location = {Sapporo, Hokkaido, Japan},
series = {HRI '22}
}

@INPROCEEDINGS{Reshapingrobot,
  author={Bucker, Arthur and Figueredo, Luis and Haddadinl, Sami and Kapoor, Ashish and Ma, Shuang and Bonatti, Rogerio},
  booktitle={2022 IEEE/RSJ International Conference on Intelligent Robots and Systems (IROS)}, 
  title={Reshaping Robot Trajectories Using Natural Language Commands: A Study of Multi-Modal Data Alignment Using Transformers}, 
  year={2022},
  volume={},
  number={},
  pages={978-984},
  doi={10.1109/IROS47612.2022.9981810}}

@misc{correctingrobotplansnatural,
      title={Correcting Robot Plans with Natural Language Feedback}, 
      author={Pratyusha Sharma and Balakumar Sundaralingam and Valts Blukis and Chris Paxton and Tucker Hermans and Antonio Torralba and Jacob Andreas and Dieter Fox},
      year={2022},
      eprint={2204.05186},
      archivePrefix={arXiv},
      primaryClass={cs.RO},
      url={https://arxiv.org/abs/2204.05186}, 
}

@INPROCEEDINGS{bringinganatural,
  author={Li, Chen and Hansen, Andreas Kornmaaler and Chrysostomou, Dimitrios and Bøgh, Simon and Madsen, Ole},
  booktitle={2022 IEEE/SICE International Symposium on System Integration (SII)}, 
  title={Bringing a Natural Language-enabled Virtual Assistant to Industrial Mobile Robots for Learning, Training and Assistance of Manufacturing Tasks}, 
  year={2022},
  volume={},
  number={},
  pages={238-243},
  doi={10.1109/SII52469.2022.9708757}}

@INPROCEEDINGS{EnablingRobotstoUnderstand,
  author={Chen, Haonan and Tan, Hao and Kuntz, Alan and Bansal, Mohit and Alterovitz, Ron},
  booktitle={2020 IEEE International Conference on Robotics and Automation (ICRA)}, 
  title={Enabling Robots to Understand Incomplete Natural Language Instructions Using Commonsense Reasoning}, 
  year={2020},
  volume={},
  number={},
  pages={1963-1969},
  doi={10.1109/ICRA40945.2020.9197315}}

@inbook{RecognisingFlexibleIntent,
author = {Wilcock, Graham},
year = {2022},
month = {01},
pages = {142-153},
title = {Recognising Flexible Intents and Multiple Domains in Extended Human-Robot Dialogues},
isbn = {978-3-030-96450-4},
doi = {10.1007/978-3-030-96451-1_13}
}

@inproceedings{Language-ConditionedImitation,
 author = {Stepputtis, Simon and Campbell, Joseph and Phielipp, Mariano and Lee, Stefan and Baral, Chitta and Ben Amor, Heni},
 booktitle = {Advances in Neural Information Processing Systems},
 editor = {H. Larochelle and M. Ranzato and R. Hadsell and M.F. Balcan and H. Lin},
 pages = {13139--13150},
 publisher = {Curran Associates, Inc.},
 title = {Language-Conditioned Imitation Learning for Robot Manipulation Tasks},
 url = {https://proceedings.neurips.cc/paper_files/paper/2020/file/9909794d52985cbc5d95c26e31125d1a-Paper.pdf},
 volume = {33},
 year = {2020}
}

@InProceedings{APersistentSpatialSemantic,
  title = 	 {A Persistent Spatial Semantic Representation for High-level Natural Language Instruction Execution},
  author =       {Blukis, Valts and Paxton, Chris and Fox, Dieter and Garg, Animesh and Artzi, Yoav},
  booktitle = 	 {Proceedings of the 5th Conference on Robot Learning},
  pages = 	 {706--717},
  year = 	 {2022},
  editor = 	 {Faust, Aleksandra and Hsu, David and Neumann, Gerhard},
  volume = 	 {164},
  series = 	 {Proceedings of Machine Learning Research},
  month = 	 {08--11 Nov},
  publisher =    {PMLR},
  url = 	 {https://proceedings.mlr.press/v164/blukis22a.html}
}

@article{WhatsonYourMind,
  author = "Kerzel, Matthias and Ambsdorf, Jakob and Becker, Dennis and Lu, Wenhao and Strahl, Erik and Spisak, Josua and Gäde, Connor and Weber, Tom and Wermter, Stefan",
  title = "What’s on Your Mind, NICO?",
  year = "2022",
  doi = "10.1007/s13218-022-00772-8",
  journal = "KI - Künstliche Intelligenz",
  volume = "36",
  number = "0",
  publisher = "Springer",
  pissn = "1610-1987",
  pages = "237--254"
}

@inproceedings{ExploringtheImpactofExplanation,
author = {Halilovic, Amar and Chandrayan, Vanchha and Krivic, Senka},
title = {Exploring the Impact of Explanation Representation on User Satisfaction in Robot Navigation},
year = {2024},
isbn = {9798400716614},
publisher = {Association for Computing Machinery},
address = {New York, NY, USA},
url = {https://doi.org/10.1145/3648536.3648537},
doi = {10.1145/3648536.3648537},
booktitle = {Proceedings of the 2024 International Symposium on Technological Advances in Human-Robot Interaction},
pages = {1–9},
numpages = {9},
location = {Boulder, CO, USA},
series = {TAHRI '24}
}

@inproceedings{ExplainableAIforRobotFailure,
author = {Das, Devleena and Banerjee, Siddhartha and Chernova, Sonia},
title = {Explainable AI for Robot Failures: Generating Explanations that Improve User Assistance in Fault Recovery},
year = {2021},
isbn = {9781450382892},
publisher = {Association for Computing Machinery},
address = {New York, NY, USA},
url = {https://doi.org/10.1145/3434073.3444657},
doi = {10.1145/3434073.3444657},
booktitle = {Proceedings of the 2021 ACM/IEEE International Conference on Human-Robot Interaction},
pages = {351–360},
numpages = {10},
location = {Boulder, CO, USA},
series = {HRI '21}
}

@inproceedings{ConversationalLanguageModelsfor,
author = {Hunt, William and Godfrey, Toby and Soorati, Mohammad D.},
title = {Conversational Language Models for Human-in-the-Loop Multi-Robot Coordination},
year = {2024},
isbn = {9798400704864},
publisher = {International Foundation for Autonomous Agents and Multiagent Systems},
address = {Richland, SC},
booktitle = {Proceedings of the 23rd International Conference on Autonomous Agents and Multiagent Systems},
pages = {2809–2811},
numpages = {3},
location = {Auckland, New Zealand},
series = {AAMAS '24}
}

@misc{SMART-LLM:SmartMulti-Agent,
      title={SMART-LLM: Smart Multi-Agent Robot Task Planning using Large Language Models}, 
      author={Shyam Sundar Kannan and Vishnunandan L. N. Venkatesh and Byung-Cheol Min},
      year={2024},
      eprint={2309.10062},
      archivePrefix={arXiv},
      primaryClass={cs.RO},
      url={https://arxiv.org/abs/2309.10062}, 
}

@inproceedings{Formalmodelingandverification,
author = {Lestingi, Livia and Sbrolli, Cristian and Scarmozzino, Pasquale and Romeo, Giorgio and Bersani, Marcello M. and Rossi, Matteo},
title = {Formal modeling and verification of multi-robot interactive scenarios in service settings},
year = {2022},
isbn = {9781450392877},
publisher = {Association for Computing Machinery},
address = {New York, NY, USA},
url = {https://doi.org/10.1145/3524482.3527653},
doi = {10.1145/3524482.3527653},
booktitle = {Proceedings of the IEEE/ACM 10th International Conference on Formal Methods in Software Engineering},
pages = {80–90},
numpages = {11},
location = {Pittsburgh, Pennsylvania},
series = {FormaliSE '22}
}

@misc{RoCo:DialecticMulti-Robo,
      title={RoCo: Dialectic Multi-Robot Collaboration with Large Language Models}, 
      author={Zhao Mandi and Shreeya Jain and Shuran Song},
      year={2023},
      eprint={2307.04738},
      archivePrefix={arXiv},
      primaryClass={cs.RO},
      url={https://arxiv.org/abs/2307.04738}, 
}

@inproceedings{PushThatThereTabletop,
author = {Wang, Keru and Wang, Zhu and Nakagaki, Ken and Perlin, Ken},
title = {“Push-That-There”: Tabletop Multi-robot Object Manipulation via Multimodal 'Object-level Instruction'},
year = {2024},
isbn = {9798400705830},
publisher = {Association for Computing Machinery},
address = {New York, NY, USA},
url = {https://doi.org/10.1145/3643834.3661542},
doi = {10.1145/3643834.3661542},
booktitle = {Proceedings of the 2024 ACM Designing Interactive Systems Conference},
pages = {2497–2513},
numpages = {17},
location = {Copenhagen, Denmark},
series = {DIS '24}
}

@INPROCEEDINGS{HowManyRobotsDoYouWant?,
  author={Zhang, Mengni and Xu, Tong and Hardin, Jackson and Jiaqi Cai, Jilly and Brooks, Johnell and Green, Keith E.},
  booktitle={2021 30th IEEE International Conference on Robot \& Human Interactive Communication (RO-MAN)}, 
  title={How Many Robots Do You Want? A Cross-Cultural Exploration on User Preference and Perception of an Assistive Multi-Robot System}, 
  year={2021},
  volume={},
  number={},
  pages={580-585},
  doi={10.1109/RO-MAN50785.2021.9515396}}

@Article{AMultirobotSysteminanAssisted,
AUTHOR = {Barber, Ramón and Ortiz, Francisco J. and Garrido, Santiago and Calatrava-Nicolás, Francisco M. and Mora, Alicia and Prados, Adrián and Vera-Repullo, José Alfonso and Roca-González, Joaquín and Méndez, Inmaculada and Mozos, Óscar Martínez},
TITLE = {A Multirobot System in an Assisted Home Environment to Support the Elderly in Their Daily Lives},
JOURNAL = {Sensors},
VOLUME = {22},
YEAR = {2022},
NUMBER = {20},
ARTICLE-NUMBER = {7983},
URL = {https://www.mdpi.com/1424-8220/22/20/7983},
PubMedID = {36298332},
ISSN = {1424-8220},
DOI = {10.3390/s22207983}
}

@inproceedings{ASmartHomeBased,
author = {Zhang, Tianqi and Zhao, Donghui and Yang, Junyou and Wang, Shuoyu and Liu, Houde},
title = {A Smart Home Based on Multi-heterogeneous Robots and Sensor Networks for Elderly Care},
year = {2022},
isbn = {978-3-031-13843-0},
publisher = {Springer-Verlag},
address = {Berlin, Heidelberg},
url = {https://doi.org/10.1007/978-3-031-13844-7_10},
doi = {10.1007/978-3-031-13844-7_10},
booktitle = {Intelligent Robotics and Applications: 15th International Conference, ICIRA 2022, Harbin, China, August 1–3, 2022, Proceedings, Part I},
pages = {98–104},
numpages = {7},
location = {Harbin, China}
}

@ARTICLE{AMultimodalEmotionalHuman–Robot,
  author={Hong, Alexander and Lunscher, Nolan and Hu, Tianhao and Tsuboi, Yuma and Zhang, Xinyi and Franco dos Reis Alves, Silas and Nejat, Goldie and Benhabib, Beno},
  journal={IEEE Transactions on Cybernetics}, 
  title={A Multimodal Emotional Human–Robot Interaction Architecture for Social Robots Engaged in Bidirectional Communication}, 
  year={2021},
  volume={51},
  number={12},
  pages={5954-5968},
  doi={10.1109/TCYB.2020.2974688}}

@article{AHumanoidSocialRobotBased,
author = {Ribino, Patrizia and Bonomolo, Marina and Lodato, Carmelo and Vitale, G.},
year = {2021},
month = {04},
pages = {},
title = {A Humanoid Social Robot Based Approach for Indoor Environment Quality Monitoring and Well-Being Improvement},
volume = {13},
journal = {International Journal of Social Robotics},
doi = {10.1007/s12369-020-00638-9}
}

@article{IntegratingSocialAssistiveRobots,
author = {Luperto, Matteo and Monroy, Javier and Renoux, Jennifer and Lunardini, Francesca and Basilico, Nicola and Bulgheroni, Maria and Cangelosi, Angelo and Cesari, Matteo and Cid, Manuel and Ianes, Aladar and González-Jiménez, Javier and Kounoudes, Anastasis and Mari, David and Prisacariu, Victor and Savanovic, Arso and Ferrante, Simona and Borghese, Alberto},
year = {2022},
month = {02},
pages = {1-31},
title = {Integrating Social Assistive Robots, IoT, Virtual Communities and Smart Objects to Assist at-Home Independently Living Elders: the MoveCare Project},
volume = {15},
journal = {International Journal of Social Robotics},
doi = {10.1007/s12369-021-00843-0}
}

@article{Mini:ANewSocialRobot,
author = {Salichs, Miguel and Castro-González, Álvaro and Salichs, Esther and Fernández-Rodicio, Enrique and Maroto Gómez, Marcos and Gamboa, Juan José and Marques, Sara and Castillo, José and Alonso-Martín, Fernando and Malfaz, Maria},
year = {2020},
month = {12},
pages = {},
title = {Mini: A New Social Robot for the Elderly},
volume = {12},
journal = {International Journal of Social Robotics},
doi = {10.1007/s12369-020-00687-0}
}

@inbook{ApplicationofRoboticstoDomestic,
author = {Ananthanarayanan, Amritha and Frazelle, Chase and Kethireddy, Sowmya and Ko, Chen-Ho and Kumar, Rohan and Prabhu, Vignesh and Vasudevan, Bhargav and Walker, Ian},
year = {2022},
month = {01},
pages = {657-665},
title = {Application of Robotics to Domestic and Environmental Cleanup Tasks},
isbn = {978-3-030-80118-2},
doi = {10.1007/978-3-030-80119-9_42}
}

@ARTICLE{ASurveyonTechniquesandApplications,
  author={Li, Zhenjing and Xu, Qingsong and Tam, Lap Mou},
  journal={IEEE Access}, 
  title={A Survey on Techniques and Applications of Window-Cleaning Robots}, 
  year={2021},
  volume={9},
  number={},
  pages={111518-111532},
  doi={10.1109/ACCESS.2021.3103757}}

@Article{SmartCleaner:ANewAutonomous,
AUTHOR = {Ruan, Kaicheng and Wu, Zehao and Xu, Qingsong},
TITLE = {Smart Cleaner: A New Autonomous Indoor Disinfection Robot for Combating the COVID-19 Pandemic},
JOURNAL = {Robotics},
VOLUME = {10},
YEAR = {2021},
NUMBER = {3},
ARTICLE-NUMBER = {87},
URL = {https://www.mdpi.com/2218-6581/10/3/87},
ISSN = {2218-6581},
DOI = {10.3390/robotics10030087}
}

@ARTICLE{AutonomousSelf-ReconfigurableFloor,
  author={Parween, Rizuwana and Vega Heredia, Manuel and Rayguru, Madan Mohan and Enjikalayil Abdulkader, Raihan and Elara, Mohan Rajesh},
  journal={IEEE Access}, 
  title={Autonomous Self-Reconfigurable Floor Cleaning Robot}, 
  year={2020},
  volume={8},
  number={},
  pages={114433-114442},
  doi={10.1109/ACCESS.2020.2999202}}

@Article{TableCleaningTask,
AUTHOR = {Yin, Jia and Apuroop, Koppaka Ganesh Sai and Tamilselvam, Yokhesh Krishnasamy and Mohan, Rajesh Elara and Ramalingam, Balakrishnan and Le, Anh Vu},
TITLE = {Table Cleaning Task by Human Support Robot Using Deep Learning Technique},
JOURNAL = {Sensors},
VOLUME = {20},
YEAR = {2020},
NUMBER = {6},
ARTICLE-NUMBER = {1698},
URL = {https://www.mdpi.com/1424-8220/20/6/1698},
PubMedID = {32197483},
ISSN = {1424-8220},
DOI = {10.3390/s20061698}
}

@article{EEG-ControlledWall-Crawling,
author = {Shao, Lei and Zhang, Longyu and Belkacem, Abdelkader Nasreddine and Zhang, Yiming and Chen, Xiaoqi and Li, Ji and Liu, Hongli},
title = {EEG-Controlled Wall-Crawling Cleaning Robot Using SSVEP-Based Brain-Computer Interface},
journal = {Journal of Healthcare Engineering},
volume = {2020},
number = {1},
pages = {6968713},
doi = {https://doi.org/10.1155/2020/6968713},
url = {https://onlinelibrary.wiley.com/doi/abs/10.1155/2020/6968713},
eprint = {https://onlinelibrary.wiley.com/doi/pdf/10.1155/2020/6968713},
year = {2020}
}

@Article{AHumanSupportRobotforthe,
AUTHOR = {Ramalingam, Balakrishnan and Yin, Jia and Rajesh Elara, Mohan and Tamilselvam, Yokhesh Krishnasamy and Mohan Rayguru, Madan and Muthugala, M. A. Viraj J. and Félix Gómez, Braulio},
TITLE = {A Human Support Robot for the Cleaning and Maintenance of Door Handles Using a Deep-Learning Framework},
JOURNAL = {Sensors},
VOLUME = {20},
YEAR = {2020},
NUMBER = {12},
ARTICLE-NUMBER = {3543},
URL = {https://www.mdpi.com/1424-8220/20/12/3543},
PubMedID = {32585864},
ISSN = {1424-8220},
DOI = {10.3390/s20123543}
}

@ARTICLE{ASwitchableUnmanned,
  author={Sun, Yinshuai and Jing, Zhongliang and Dong, Peng and Huang, Jianzhe and Chen, Wujun and Leung, Henry},
  journal={IEEE Robotics and Automation Letters}, 
  title={A Switchable Unmanned Aerial Manipulator System for Window-Cleaning Robot Installation}, 
  year={2021},
  volume={6},
  number={2},
  pages={3483-3490},
  doi={10.1109/LRA.2021.3062795}}

@article{sTetro-D,
title = {sTetro-D: A deep learning based autonomous descending-stair cleaning robot},
journal = {Engineering Applications of Artificial Intelligence},
volume = {120},
pages = {105844},
year = {2023},
issn = {0952-1976},
doi = {https://doi.org/10.1016/j.engappai.2023.105844},
url = {https://www.sciencedirect.com/science/article/pii/S0952197623000283},
author = {Veerajagadheswar Prabakaran and Anh Vu Le and Phone Thiha Kyaw and Prathap Kandasamy and Aung Paing and Rajesh Elara Mohan}
}

@Article{TowardsIndoorSuctionable,
AUTHOR = {Huang, Qian},
TITLE = {Towards Indoor Suctionable Object Classification and Recycling: Developing a Lightweight AI Model for Robot Vacuum Cleaners},
JOURNAL = {Applied Sciences},
VOLUME = {13},
YEAR = {2023},
NUMBER = {18},
ARTICLE-NUMBER = {10031},
URL = {https://www.mdpi.com/2076-3417/13/18/10031},
ISSN = {2076-3417},
DOI = {10.3390/app131810031}
}

@article{ENRICHME:Perception,
author = {Coşar, Serhan and Fernández-Carmona, Manuel and Agrigoroaie, Roxana and Pages, Jordi and Ferland, François and Zhao, Feng and Yue, Shigang and Bellotto, Nicola and Tapus, Adriana},
year = {2020},
month = {07},
pages = {},
title = {ENRICHME: Perception and Interaction of an Assistive Robot for the Elderly at Home},
volume = {12},
journal = {International Journal of Social Robotics},
doi = {10.1007/s12369-019-00614-y}
}

@Article{Home-BasedRehabilitationof,
AUTHOR = {Cunha, Bruno and Ferreira, Ricardo and Sousa, Andreia S. P.},
TITLE = {Home-Based Rehabilitation of the Shoulder Using Auxiliary Systems and Artificial Intelligence: An Overview},
JOURNAL = {Sensors},
VOLUME = {23},
YEAR = {2023},
NUMBER = {16},
ARTICLE-NUMBER = {7100},
URL = {https://www.mdpi.com/1424-8220/23/16/7100},
PubMedID = {37631637},
ISSN = {1424-8220},
DOI = {10.3390/s23167100}
}

@article{Utilizingtheintelligenceedge,
title = {Utilizing the intelligence edge framework for robotic upper limb rehabilitation in home},
journal = {MethodsX},
volume = {11},
pages = {102312},
year = {2023},
issn = {2215-0161},
doi = {https://doi.org/10.1016/j.mex.2023.102312},
url = {https://www.sciencedirect.com/science/article/pii/S2215016123003096},
author = {Prashant K. Jamwal and Aibek Niyetkaliyev and Shahid Hussain and Aditi Sharma and Paulette {Van Vliet}}
}

@inproceedings{DesigningParent-child-robot,
author = {Ho, Hui-Ru and White, Nathan Thomas and Hubbard, Edward M. and Mutlu, Bilge},
title = {Designing Parent-child-robot Interactions to Facilitate In-Home Parental Math Talk with Young Children},
year = {2023},
isbn = {9798400701313},
publisher = {Association for Computing Machinery},
address = {New York, NY, USA},
url = {https://doi.org/10.1145/3585088.3589358},
doi = {10.1145/3585088.3589358},
booktitle = {Proceedings of the 22nd Annual ACM Interaction Design and Children Conference},
pages = {355–366},
numpages = {12},
location = {Chicago, IL, USA},
series = {IDC '23}
}

@inproceedings{RobocampatHome:Exploring,
author = {Ahtinen, Aino and Beheshtian, Nasim and V\"{a}\"{a}n\"{a}nen, Kaisa},
title = {Robocamp at Home: Exploring Families' Co-Learning with a Social Robot: Findings from a One-Month Study in the Wild},
year = {2023},
isbn = {9781450399647},
publisher = {Association for Computing Machinery},
address = {New York, NY, USA},
url = {https://doi.org/10.1145/3568162.3576976},
doi = {10.1145/3568162.3576976},
booktitle = {Proceedings of the 2023 ACM/IEEE International Conference on Human-Robot Interaction},
pages = {331–340},
numpages = {10},
location = {Stockholm, Sweden},
series = {HRI '23}
}

@article{Robottutorandpupils’educational,
title = {Robot tutor and pupils’ educational ability: Teaching the times tables},
journal = {Computers \& Education},
volume = {157},
pages = {103970},
year = {2020},
issn = {0360-1315},
doi = {https://doi.org/10.1016/j.compedu.2020.103970},
url = {https://www.sciencedirect.com/science/article/pii/S0360131520301688},
author = {Elly A. Konijn and Johan F. Hoorn}
}

@article{socialrobotforlongterm,
author = {Leite, Iolanda and Martinho, Carlos and Paiva, Ana},
year = {2013},
month = {04},
pages = {},
title = {Social Robots for Long-Term Interaction: A Survey},
volume = {5},
journal = {International Journal of Social Robotics},
doi = {10.1007/s12369-013-0178-y}
}

@Inbook{socialrobot,
author="Breazeal, Cynthia
and Dautenhahn, Kerstin
and Kanda, Takayuki",
editor="Siciliano, Bruno
and Khatib, Oussama",
title="Social Robotics",
bookTitle="Springer Handbook of Robotics",
year="2016",
publisher="Springer International Publishing",
address="Cham",
pages="1935--1972",
isbn="978-3-319-32552-1",
doi="10.1007/978-3-319-32552-1_72",
url="https://doi.org/10.1007/978-3-319-32552-1_72"
}

@article{alexa,
author = {Irene Lopatovska and Katrina Rink and Ian Knight and Kieran Raines and Kevin Cosenza and Harriet Williams and Perachya Sorsche and David Hirsch and Qi Li and Adrianna Martinez},
title ={Talk to me: Exploring user interactions with the Amazon Alexa},
journal = {Journal of Librarianship and Information Science},
volume = {51},
number = {4},
pages = {984-997},
year = {2019},
doi = {10.1177/0961000618759414},
URL = { 
        https://doi.org/10.1177/0961000618759414
},
eprint = { 
        https://doi.org/10.1177/0961000618759414
}
}

@misc{GoogleAssistant,
  author = {{Google}},
  title = {Google Assistant},
  year = {2016},
  howpublished = {\url{https://assistant.google.com/}},
  note = {Accessed: 2024-08-22}
}

@misc{AppleSiri,
  author = {{Apple Inc.}},
  title = {Siri},
  year = {2011},
  howpublished = {\url{https://www.apple.com/siri/}},
  note = {Accessed: 2024-08-22}
}

@misc{openai2024gpt4technicalreport,
      title={GPT-4 Technical Report}, 
      author={OpenAI and Josh Achiam and Steven Adler and Sandhini Agarwal and Lama Ahmad and Ilge Akkaya and Florencia Leoni Aleman and Diogo Almeida and Janko Altenschmidt and Sam Altman and Shyamal Anadkat and Red Avila and Igor Babuschkin and Suchir Balaji and Valerie Balcom and Paul Baltescu and Haiming Bao and Mohammad Bavarian and Jeff Belgum and Irwan Bello and Jake Berdine and Gabriel Bernadett-Shapiro and Christopher Berner and Lenny Bogdonoff and Oleg Boiko and Madelaine Boyd and Anna-Luisa Brakman and Greg Brockman and Tim Brooks and Miles Brundage and Kevin Button and Trevor Cai and Rosie Campbell and Andrew Cann and Brittany Carey and Chelsea Carlson and Rory Carmichael and Brooke Chan and Che Chang and Fotis Chantzis and Derek Chen and Sully Chen and Ruby Chen and Jason Chen and Mark Chen and Ben Chess and Chester Cho and Casey Chu and Hyung Won Chung and Dave Cummings and Jeremiah Currier and Yunxing Dai and Cory Decareaux and Thomas Degry and Noah Deutsch and Damien Deville and Arka Dhar and David Dohan and Steve Dowling and Sheila Dunning and Adrien Ecoffet and Atty Eleti and Tyna Eloundou and David Farhi and Liam Fedus and Niko Felix and Simón Posada Fishman and Juston Forte and Isabella Fulford and Leo Gao and Elie Georges and Christian Gibson and Vik Goel and Tarun Gogineni and Gabriel Goh and Rapha Gontijo-Lopes and Jonathan Gordon and Morgan Grafstein and Scott Gray and Ryan Greene and Joshua Gross and Shixiang Shane Gu and Yufei Guo and Chris Hallacy and Jesse Han and Jeff Harris and Yuchen He and Mike Heaton and Johannes Heidecke and Chris Hesse and Alan Hickey and Wade Hickey and Peter Hoeschele and Brandon Houghton and Kenny Hsu and Shengli Hu and Xin Hu and Joost Huizinga and Shantanu Jain and Shawn Jain and Joanne Jang and Angela Jiang and Roger Jiang and Haozhun Jin and Denny Jin and Shino Jomoto and Billie Jonn and Heewoo Jun and Tomer Kaftan and Łukasz Kaiser and Ali Kamali and Ingmar Kanitscheider and Nitish Shirish Keskar and Tabarak Khan and Logan Kilpatrick and Jong Wook Kim and Christina Kim and Yongjik Kim and Jan Hendrik Kirchner and Jamie Kiros and Matt Knight and Daniel Kokotajlo and Łukasz Kondraciuk and Andrew Kondrich and Aris Konstantinidis and Kyle Kosic and Gretchen Krueger and Vishal Kuo and Michael Lampe and Ikai Lan and Teddy Lee and Jan Leike and Jade Leung and Daniel Levy and Chak Ming Li and Rachel Lim and Molly Lin and Stephanie Lin and Mateusz Litwin and Theresa Lopez and Ryan Lowe and Patricia Lue and Anna Makanju and Kim Malfacini and Sam Manning and Todor Markov and Yaniv Markovski and Bianca Martin and Katie Mayer and Andrew Mayne and Bob McGrew and Scott Mayer McKinney and Christine McLeavey and Paul McMillan and Jake McNeil and David Medina and Aalok Mehta and Jacob Menick and Luke Metz and Andrey Mishchenko and Pamela Mishkin and Vinnie Monaco and Evan Morikawa and Daniel Mossing and Tong Mu and Mira Murati and Oleg Murk and David Mély and Ashvin Nair and Reiichiro Nakano and Rajeev Nayak and Arvind Neelakantan and Richard Ngo and Hyeonwoo Noh and Long Ouyang and Cullen O'Keefe and Jakub Pachocki and Alex Paino and Joe Palermo and Ashley Pantuliano and Giambattista Parascandolo and Joel Parish and Emy Parparita and Alex Passos and Mikhail Pavlov and Andrew Peng and Adam Perelman and Filipe de Avila Belbute Peres and Michael Petrov and Henrique Ponde de Oliveira Pinto and Michael and Pokorny and Michelle Pokrass and Vitchyr H. Pong and Tolly Powell and Alethea Power and Boris Power and Elizabeth Proehl and Raul Puri and Alec Radford and Jack Rae and Aditya Ramesh and Cameron Raymond and Francis Real and Kendra Rimbach and Carl Ross and Bob Rotsted and Henri Roussez and Nick Ryder and Mario Saltarelli and Ted Sanders and Shibani Santurkar and Girish Sastry and Heather Schmidt and David Schnurr and John Schulman and Daniel Selsam and Kyla Sheppard and Toki Sherbakov and Jessica Shieh and Sarah Shoker and Pranav Shyam and Szymon Sidor and Eric Sigler and Maddie Simens and Jordan Sitkin and Katarina Slama and Ian Sohl and Benjamin Sokolowsky and Yang Song and Natalie Staudacher and Felipe Petroski Such and Natalie Summers and Ilya Sutskever and Jie Tang and Nikolas Tezak and Madeleine B. Thompson and Phil Tillet and Amin Tootoonchian and Elizabeth Tseng and Preston Tuggle and Nick Turley and Jerry Tworek and Juan Felipe Cerón Uribe and Andrea Vallone and Arun Vijayvergiya and Chelsea Voss and Carroll Wainwright and Justin Jay Wang and Alvin Wang and Ben Wang and Jonathan Ward and Jason Wei and CJ Weinmann and Akila Welihinda and Peter Welinder and Jiayi Weng and Lilian Weng and Matt Wiethoff and Dave Willner and Clemens Winter and Samuel Wolrich and Hannah Wong and Lauren Workman and Sherwin Wu and Jeff Wu and Michael Wu and Kai Xiao and Tao Xu and Sarah Yoo and Kevin Yu and Qiming Yuan and Wojciech Zaremba and Rowan Zellers and Chong Zhang and Marvin Zhang and Shengjia Zhao and Tianhao Zheng and Juntang Zhuang and William Zhuk and Barret Zoph},
      year={2024},
      eprint={2303.08774},
      archivePrefix={arXiv},
      primaryClass={cs.CL},
      url={https://arxiv.org/abs/2303.08774}, 
}

@misc{palm2,
      title={PaLM 2 Technical Report}, 
      author={Rohan Anil and Andrew M. Dai and Orhan Firat and Melvin Johnson and Dmitry Lepikhin and Alexandre Passos and Siamak Shakeri and Emanuel Taropa and Paige Bailey and Zhifeng Chen and Eric Chu and Jonathan H. Clark and Laurent El Shafey and Yanping Huang and Kathy Meier-Hellstern and Gaurav Mishra and Erica Moreira and Mark Omernick and Kevin Robinson and Sebastian Ruder and Yi Tay and Kefan Xiao and Yuanzhong Xu and Yujing Zhang and Gustavo Hernandez Abrego and Junwhan Ahn and Jacob Austin and Paul Barham and Jan Botha and James Bradbury and Siddhartha Brahma and Kevin Brooks and Michele Catasta and Yong Cheng and Colin Cherry and Christopher A. Choquette-Choo and Aakanksha Chowdhery and Clément Crepy and Shachi Dave and Mostafa Dehghani and Sunipa Dev and Jacob Devlin and Mark Díaz and Nan Du and Ethan Dyer and Vlad Feinberg and Fangxiaoyu Feng and Vlad Fienber and Markus Freitag and Xavier Garcia and Sebastian Gehrmann and Lucas Gonzalez and Guy Gur-Ari and Steven Hand and Hadi Hashemi and Le Hou and Joshua Howland and Andrea Hu and Jeffrey Hui and Jeremy Hurwitz and Michael Isard and Abe Ittycheriah and Matthew Jagielski and Wenhao Jia and Kathleen Kenealy and Maxim Krikun and Sneha Kudugunta and Chang Lan and Katherine Lee and Benjamin Lee and Eric Li and Music Li and Wei Li and YaGuang Li and Jian Li and Hyeontaek Lim and Hanzhao Lin and Zhongtao Liu and Frederick Liu and Marcello Maggioni and Aroma Mahendru and Joshua Maynez and Vedant Misra and Maysam Moussalem and Zachary Nado and John Nham and Eric Ni and Andrew Nystrom and Alicia Parrish and Marie Pellat and Martin Polacek and Alex Polozov and Reiner Pope and Siyuan Qiao and Emily Reif and Bryan Richter and Parker Riley and Alex Castro Ros and Aurko Roy and Brennan Saeta and Rajkumar Samuel and Renee Shelby and Ambrose Slone and Daniel Smilkov and David R. So and Daniel Sohn and Simon Tokumine and Dasha Valter and Vijay Vasudevan and Kiran Vodrahalli and Xuezhi Wang and Pidong Wang and Zirui Wang and Tao Wang and John Wieting and Yuhuai Wu and Kelvin Xu and Yunhan Xu and Linting Xue and Pengcheng Yin and Jiahui Yu and Qiao Zhang and Steven Zheng and Ce Zheng and Weikang Zhou and Denny Zhou and Slav Petrov and Yonghui Wu},
      year={2023},
      eprint={2305.10403},
      archivePrefix={arXiv},
      primaryClass={cs.CL},
      url={https://arxiv.org/abs/2305.10403}, 
}

@INPROCEEDINGS{designofahomemulti,
  author={Benavidez, Patrick and Kumar, Mohan and Agaian, Sos and Jamshidi, Mo},
  booktitle={2015 10th System of Systems Engineering Conference (SoSE)}, 
  title={Design of a home multi-robot system for the elderly and disabled}, 
  year={2015},
  volume={},
  number={},
  pages={392-397},
  doi={10.1109/SYSOSE.2015.7151907}}

@inproceedings{roleplay1,
author = {Matthews, Mark and Gay, Geri and Doherty, Gavin},
title = {Taking part: role-play in the design of therapeutic systems},
year = {2014},
isbn = {9781450324731},
publisher = {Association for Computing Machinery},
address = {New York, NY, USA},
url = {https://doi.org/10.1145/2556288.2557103},
doi = {10.1145/2556288.2557103},
booktitle = {Proceedings of the SIGCHI Conference on Human Factors in Computing Systems},
pages = {643–652},
numpages = {10},
location = {Toronto, Ontario, Canada},
series = {CHI '14}
}

@inproceedings{roleplay2,
  author    = {Stefan Boess},
  title     = {Rationales for Role Playing in Design},
  booktitle = {Wonderground - DRS International Conference 2006},
  editor    = {Kurt Friedman and Tom Love and Eduardo Côrte-Real and Caroline Rust},
  year      = {2006},
  pages     = {1--4},
  address   = {Lisbon, Portugal},
  url       = {https://dl.designresearchsociety.org/drs-conference-papers/drs2006/researchpapers/69},
  note      = {Accessed: 2024-08-31}
}

@article{VRoverview,
  author    = {Ahmed Hamad and Bo Jia},
  title     = {How Virtual Reality Technology Has Changed Our Lives: An Overview of the Current and Potential Applications and Limitations},
  journal   = {International Journal of Environmental Research and Public Health},
  volume    = {19},
  number    = {18},
  pages     = {11278},
  year      = {2022},
  month     = {September},
  doi       = {10.3390/ijerph191811278},
  pmid      = {36141551},
  pmcid     = {PMC9517547},
  url       = {https://www.ncbi.nlm.nih.gov/pmc/articles/PMC9517547/},
  note      = {Accessed: 2024-08-31}
}

@article{immersiveVR,
  author    = {Michael Balcerak Jackson and Barbara Balcerak Jackson},
  title     = {Immersive Experience and Virtual Reality},
  journal   = {Philosophy \& Technology},
  volume    = {37},
  number    = {1},
  pages     = {19},
  year      = {2024},
  doi       = {10.1007/s13347-024-00707-1},
  url       = {https://doi.org/10.1007/s13347-024-00707-1},
  note      = {Accessed: 2024-08-31}
}

@misc{homerobot,
      title={HomeRobot: Open-Vocabulary Mobile Manipulation}, 
      author={Sriram Yenamandra and Arun Ramachandran and Karmesh Yadav and Austin Wang and Mukul Khanna and Theophile Gervet and Tsung-Yen Yang and Vidhi Jain and Alexander William Clegg and John Turner and Zsolt Kira and Manolis Savva and Angel Chang and Devendra Singh Chaplot and Dhruv Batra and Roozbeh Mottaghi and Yonatan Bisk and Chris Paxton},
      year={2024},
      eprint={2306.11565},
      archivePrefix={arXiv},
      primaryClass={cs.RO},
      url={https://arxiv.org/abs/2306.11565}, 
}

@ARTICLE{lotenableddualarm,
  author={Zhou, Huiying and Yang, Geng and Lyu, Honghao and Huang, Xiaoyan and Yang, Huayong and Pang, Zhibo},
  journal={IEEE Journal of Biomedical and Health Informatics}, 
  title={IoT-Enabled Dual-Arm Motion Capture and Mapping for Telerobotics in Home Care}, 
  year={2020},
  volume={24},
  number={6},
  pages={1541-1549},
  doi={10.1109/JBHI.2019.2953885}}

@misc{PaLME,
      title={PaLM-E: An Embodied Multimodal Language Model}, 
      author={Danny Driess and Fei Xia and Mehdi S. M. Sajjadi and Corey Lynch and Aakanksha Chowdhery and Brian Ichter and Ayzaan Wahid and Jonathan Tompson and Quan Vuong and Tianhe Yu and Wenlong Huang and Yevgen Chebotar and Pierre Sermanet and Daniel Duckworth and Sergey Levine and Vincent Vanhoucke and Karol Hausman and Marc Toussaint and Klaus Greff and Andy Zeng and Igor Mordatch and Pete Florence},
      year={2023},
      eprint={2303.03378},
      archivePrefix={arXiv},
      primaryClass={cs.LG},
      url={https://arxiv.org/abs/2303.03378}, 
}

@inproceedings{understandingnaturallanguagecommands,
author = {Tellex, Stefanie and Kollar, Thomas and Dickerson, Steven and Walter, Matthew and Banerjee, Ashis and Teller, Seth and Roy, Nicholas},
year = {2011},
month = {01},
pages = {},
title = {Understanding Natural Language Commands for Robotic Navigation and Mobile Manipulation.},
volume = {2}
}

@inproceedings{learningtoparse,
  title={Learning to parse natural language commands to a robot control system},
  author={Matuszek, Cynthia and Herbst, Evan and Zettlemoyer, Luke and Fox, Dieter},
  booktitle={Experimental robotics: the 13th international symposium on experimental robotics},
  pages={403--415},
  year={2013},
  organization={Springer}
}

@INPROCEEDINGS{towardsunderstanding,
  author={Kollar, Thomas and Tellex, Stefanie and Roy, Deb and Roy, Nicholas},
  booktitle={2010 5th ACM/IEEE International Conference on Human-Robot Interaction (HRI)}, 
  title={Toward understanding natural language directions}, 
  year={2010},
  volume={},
  number={},
  pages={259-266},
  doi={10.1109/HRI.2010.5453186}}

@inproceedings{walkthetalk,
author = {MacMahon, Matt and Stankiewicz, Brian and Kuipers, Benjamin},
title = {Walk the talk: connecting language, knowledge, and action in route instructions},
year = {2006},
isbn = {9781577352815},
publisher = {AAAI Press},
booktitle = {Proceedings of the 21st National Conference on Artificial Intelligence - Volume 2},
pages = {1475–1482},
numpages = {8},
location = {Boston, Massachusetts},
series = {AAAI'06}
}

@article{Autonomousvacuumcleaner,
title = {Autonomous vacuum cleaner},
journal = {Robotics and Autonomous Systems},
volume = {19},
number = {3},
pages = {233-245},
year = {1997},
note = {Intelligent Robotic Systems SIRS'95},
issn = {0921-8890},
doi = {https://doi.org/10.1016/S0921-8890(96)00053-X},
url = {https://www.sciencedirect.com/science/article/pii/S092188909600053X},
author = {Iwan Ulrich and Francesco Mondada and J.-D. Nicoud}
}

@INPROCEEDINGS{EvaluatingtheRoomba,
  author={Tribelhorn, Ben and Dodds, Zachary},
  booktitle={Proceedings 2007 IEEE International Conference on Robotics and Automation}, 
  title={Evaluating the Roomba: A low-cost, ubiquitous platform for robotics research and education}, 
  year={2007},
  volume={},
  number={},
  pages={1393-1399},
  doi={10.1109/ROBOT.2007.363179}}

@article{Mobilerobotprogramming,
title = {Mobile robot programming using natural language},
journal = {Robotics and Autonomous Systems},
volume = {38},
number = {3},
pages = {171-181},
year = {2002},
note = {Advances in Robot Skill Learning},
issn = {0921-8890},
doi = {https://doi.org/10.1016/S0921-8890(02)00166-5},
url = {https://www.sciencedirect.com/science/article/pii/S0921889002001665},
author = {Stanislao Lauria and Guido Bugmann and Theocharis Kyriacou and Ewan Klein}
}

@ARTICLE{Spatiallanguagefor,
  author={Skubic, M. and Perzanowski, D. and Blisard, S. and Schultz, A. and Adams, W. and Bugajska, M. and Brock, D.},
  journal={IEEE Transactions on Systems, Man, and Cybernetics, Part C (Applications and Reviews)}, 
  title={Spatial language for human-robot dialogs}, 
  year={2004},
  volume={34},
  number={2},
  pages={154-167},
  doi={10.1109/TSMCC.2004.826273}}

@article{mobilemanipulation,
title = {Mobile manipulation: The robotic assistant},
journal = {Robotics and Autonomous Systems},
volume = {26},
number = {2},
pages = {175-183},
year = {1999},
note = {Field and Service Robotics},
issn = {0921-8890},
doi = {https://doi.org/10.1016/S0921-8890(98)00067-0},
url = {https://www.sciencedirect.com/science/article/pii/S0921889098000670},
author = {Oussama Khatib}
}

@article{Towardaframeworkforlevels,
author = {Beer, Jenay M. and Fisk, Arthur D. and Rogers, Wendy A.},
title = {Toward a framework for levels of robot autonomy in human-robot interaction},
year = {2014},
issue_date = {July 2014},
publisher = {Journal of Human-Robot Interaction Steering Committee},
volume = {3},
number = {2},
url = {https://doi.org/10.5898/JHRI.3.2.Beer},
doi = {10.5898/JHRI.3.2.Beer},
month = {jul},
pages = {74–99},
numpages = {26}
}

@inproceedings{HumanRobotMutualAdaptation, series={HRI ’17},
   title={Human-Robot Mutual Adaptation in Shared Autonomy},
   url={http://dx.doi.org/10.1145/2909824.3020252},
   DOI={10.1145/2909824.3020252},
   booktitle={Proceedings of the 2017 ACM/IEEE International Conference on Human-Robot Interaction},
   publisher={ACM},
   author={Nikolaidis, Stefanos and Zhu, Yu Xiang and Hsu, David and Srinivasa, Siddhartha},
   year={2017},
   month=mar, collection={HRI ’17} }

@article{ataxonomyoffactors,
author = {Akalin, Neziha and Kiselev, Andrey and Kristoffersson, Annica and Loutfi, Amy},
year = {2023},
month = {07},
pages = {},
title = {A Taxonomy of Factors Influencing Perceived Safety in Human–Robot Interaction},
volume = {15},
journal = {International Journal of Social Robotics},
doi = {10.1007/s12369-023-01027-8}
}

@INPROCEEDINGS{AdaptiveLeaderFollower,
  author={van Zoelen, Emma M. and Barakova, Emilia I. and Rauterberg, Matthias},
  booktitle={2020 29th IEEE International Conference on Robot and Human Interactive Communication (RO-MAN)}, 
  title={Adaptive Leader-Follower Behavior in Human-Robot Collaboration}, 
  year={2020},
  volume={},
  number={},
  pages={1259-1265},
  doi={10.1109/RO-MAN47096.2020.9223548}}

@article{PreferredInteractionStyles,
author = {Ruth Schulz and Philipp Kratzer and Marc Toussaint},
title = {Preferred Interaction Styles for Human-Robot Collaboration Vary Over Tasks With Different Action Types},
journal = {Frontiers in Neurorobotics},
year = {2018},
volume = {12},
publisher = {Frontiers Media S.A.},
month = {jul},
url = {https://doi.org/10.3389/fnbot.2018.00036},
doi = {10.3389/fnbot.2018.00036}
}

@article{RobotleadershipInvestigating,
    doi = {10.1371/journal.pone.0281786},
    author = {Cichor, Jakub Edward AND Hubner-Benz, Sylvia AND Benz, Tobias AND Emmerling, Franziska AND Peus, Claudia},
    journal = {PLOS ONE},
    publisher = {Public Library of Science},
    title = {Robot leadership–Investigating human perceptions and reactions towards social robots showing leadership behaviors},
    year = {2023},
    month = {02},
    volume = {18},
    url = {https://doi.org/10.1371/journal.pone.0281786},
    pages = {1-15},
    number = {2},

}

@article{DecisionMakingAuthority,
author = {Gombolay, Matthew and Gutierrez, Reymundo and Clarke, Shanelle and Sturla, Giancarlo and Shah, Julie},
year = {2015},
month = {07},
pages = {},
title = {Decision-Making Authority, Team Efficiency and Human Worker Satisfaction in Mixed Human-Robot Teams},
volume = {39},
journal = {Autonomous Robots},
doi = {10.1007/s10514-015-9457-9}
}

@inproceedings{InfluencingLeadingandFollowing,
author = {Kwon, Minae and Li, Mengxi and Bucquet, Alexandre and Sadigh, Dorsa},
year = {2019},
month = {06},
pages = {},
title = {Influencing Leading and Following in Human-Robot Teams},
doi = {10.15607/RSS.2019.XV.075}
}

@ARTICLE{Collaboratingeyetoeye,

AUTHOR={Arntz, Alexander  and Straßmann, Carolin  and Völker, Stefanie  and Eimler, Sabrina C. },

TITLE={Collaborating eye to eye: Effects of workplace design on the perception of dominance of collaboration robots},

JOURNAL={Frontiers in Robotics and AI},

VOLUME={9},

YEAR={2022},

URL={https://www.frontiersin.org/journals/robotics-and-ai/articles/10.3389/frobt.2022.999308},

DOI={10.3389/frobt.2022.999308},

ISSN={2296-9144}}

@article{TheWorldisnotEnough,
author = {de Visser, Ewart and Krueger, Frank and Mcknight, Patrick and Scheid, Steven and Smith, Melissa and Chalk, Stephanie and Parasuraman, Raja},
year = {2012},
month = {10},
pages = {263-267},
title = {The World is not Enough: Trust in Cognitive Agents},
volume = {56},
journal = {Proceedings of the Human Factors and Ergonomics Society Annual Meeting},
doi = {10.1177/1071181312561062}
}

@article{Adaptiveassistiverobotics,
author = {Gordon, Daniel and Christou, Andreas and Stouraitis, Theodoros and Gienger, Michael and Vijayakumar, Sethu},
year = {2023},
month = {06},
pages = {},
title = {Adaptive assistive robotics: a framework for triadic collaboration between humans and robots},
volume = {10},
journal = {Royal Society Open Science},
doi = {10.1098/rsos.221617}
}

@article{Theeffectofcognitive,
author = {Karakikes, Michail and Nathanael, Dimitris},
year = {2022},
month = {11},
pages = {1-13},
title = {The effect of cognitive workload on decision authority assignment in human–robot collaboration},
volume = {25},
journal = {Cognition, Technology \& Work},
doi = {10.1007/s10111-022-00719-x}
}

@article{Humanrobotmutual,
author = {Nikolaidis, Stefanos and Hsu, David and Srinivasa, Siddhartha},
title = {Human-robot mutual adaptation in collaborative tasks},
year = {2017},
issue_date = {6 2017},
publisher = {Sage Publications, Inc.},
address = {USA},
volume = {36},
number = {5–7},
issn = {0278-3649},
url = {https://doi.org/10.1177/0278364917690593},
doi = {10.1177/0278364917690593},
journal = {Int. J. Rob. Res.},
month = {jun},
pages = {618–634},
numpages = {17}
}

@article{HumanPerformanceIssues,
author = {Chen, Jessie and Haas, Ellen and Barnes, Michael},
year = {2007},
month = {12},
pages = {1231 - 1245},
title = {Human Performance Issues and User Interface Design for Teleoperated Robots},
volume = {37},
journal = {Systems, Man, and Cybernetics, Part C: Applications and Reviews, IEEE Transactions on},
doi = {10.1109/TSMCC.2007.905819}
}

@book{Teleroboticsautomation,
author = {Sheridan, Thomas B.},
title = {Telerobotics, automation, and human supervisory control},
year = {1992},
isbn = {0262193167},
publisher = {MIT Press},
address = {Cambridge, MA, USA}
}

@article{TrustAwareDecisionMaking,
author = {Chen, Min and Nikolaidis, Stefanos and Soh, Harold and Hsu, David and Srinivasa, Siddhartha},
title = {Trust-Aware Decision Making for Human-Robot Collaboration: Model Learning and Planning},
year = {2020},
issue_date = {June 2020},
publisher = {Association for Computing Machinery},
address = {New York, NY, USA},
volume = {9},
number = {2},
url = {https://doi.org/10.1145/3359616},
doi = {10.1145/3359616},
journal = {J. Hum.-Robot Interact.},
month = {jan},
articleno = {9},
numpages = {23}
}

@inproceedings{thetransferofhuman,
author = {Soh, Harold and Pan, Shu and Min, Chen and Hsu, David},
year = {2018},
month = {06},
pages = {},
title = {The Transfer of Human Trust in Robot Capabilities across Tasks},
doi = {10.15607/RSS.2018.XIV.033}
}

@article{ASurveyonTrustinAutonomous,
  title={A Survey on Trust in Autonomous Systems},
  author={Shervin Shahrdar and Luiza Menezes and Mehrdad Nojoumian},
  journal={Advances in Intelligent Systems and Computing},
  year={2018},
  url={https://api.semanticscholar.org/CorpusID:3737342}
}

@article{AnthropomorphismOpportunities,
  title={Anthropomorphism: Opportunities and Challenges in Human–Robot Interaction},
  author={Jakub Złotowski and Diane Proudfoot and Kumar Yogeeswaran and Christoph Bartneck},
  journal={International Journal of Social Robotics},
  year={2014},
  volume={7},
  pages={347 - 360},
  url={https://api.semanticscholar.org/CorpusID:4832420}
}

@article{Anthropomorphisminhumanrobot,
    author = {Kühne, Rinaldo and Peter, Jochen},
    title = "{Anthropomorphism in human–robot interactions: a multidimensional conceptualization}",
    journal = {Communication Theory},
    volume = {33},
    number = {1},
    pages = {42-52},
    year = {2022},
    month = {10},
    issn = {1468-2885},
    doi = {10.1093/ct/qtac020},
    url = {https://doi.org/10.1093/ct/qtac020},
    eprint = {https://academic.oup.com/ct/article-pdf/33/1/42/48958537/qtac020.pdf},
}

@article{Asurveyofsocially,
title = {A survey of socially interactive robots},
journal = {Robotics and Autonomous Systems},
volume = {42},
number = {3},
pages = {143-166},
year = {2003},
note = {Socially Interactive Robots},
issn = {0921-8890},
doi = {https://doi.org/10.1016/S0921-8890(02)00372-X},
url = {https://www.sciencedirect.com/science/article/pii/S092188900200372X},
author = {Terrence Fong and Illah Nourbakhsh and Kerstin Dautenhahn},
}

@article{Towardsociablerobots,
title = {Toward sociable robots},
journal = {Robotics and Autonomous Systems},
volume = {42},
number = {3},
pages = {167-175},
year = {2003},
note = {Socially Interactive Robots},
issn = {0921-8890},
doi = {https://doi.org/10.1016/S0921-8890(02)00373-1},
url = {https://www.sciencedirect.com/science/article/pii/S0921889002003731},
author = {Cynthia Breazeal},
}

@INPROCEEDINGS{Humanoidrobotsasa,
  author={Hayashi, Kotaro and Sakamoto, Daisuke and Kanda, Takayuki and Shiomi, Masahiro and Koizumi, Satoshi and Ishiguro, Hiroshi and Ogasawara, Tsukasa and Hagita, Norihiro},
  booktitle={2007 2nd ACM/IEEE International Conference on Human-Robot Interaction (HRI)}, 
  title={Humanoid robots as a passive-social medium - a field experiment at a train station}, 
  year={2007},
  volume={},
  number={},
  pages={137-144},
  doi={10.1145/1228716.1228735}}

@article{TowardsanIntegrative,
author = {Dobrosovestnova, Anna and Reinboth, Tim and Weiss, Astrid},
title = {Towards an Integrative Framework for Robot Personality Research},
year = {2024},
issue_date = {March 2024},
publisher = {Association for Computing Machinery},
address = {New York, NY, USA},
volume = {13},
number = {1},
url = {https://doi.org/10.1145/3640010},
doi = {10.1145/3640010},
month = {feb},
articleno = {8},
numpages = {22}
}

@inproceedings{PersonalityintheHumanRobot,
author = {Robert, Lionel},
year = {2018},
month = {05},
pages = {},
title = {Personality in the Human Robot Interaction Literature: A Review and Brief Critique}
}

@article{ThreeResponsesto,
  title={Three Responses to Anthropomorphism in Social Robotics: Towards a Critical, Relational, and Hermeneutic Approach},
  author={Mark Coeckelbergh},
  journal={International Journal of Social Robotics},
  year={2021},
  volume={14},
  pages={2049 - 2061},
  url={https://api.semanticscholar.org/CorpusID:233703071}
}

@misc{huang2024rekepspatiotemporalreasoningrelational,
      title={ReKep: Spatio-Temporal Reasoning of Relational Keypoint Constraints for Robotic Manipulation}, 
      author={Wenlong Huang and Chen Wang and Yunzhu Li and Ruohan Zhang and Li Fei-Fei},
      year={2024},
      eprint={2409.01652},
      archivePrefix={arXiv},
      primaryClass={cs.RO},
      url={https://arxiv.org/abs/2409.01652}, 
}

@misc{amazonastro2024,
  title = {Introducing Amazon Astro},
  author={Amazon},
  howpublished = {\url{https://www.amazon.com/Introducing-Amazon-Astro/dp/B078NSDFSB}},
  year = {2024},
  note = {Accessed: 2024-09-11}
}

@misc{meetobi2024,
  title = {Meet Obi},
  author={Obi},
  howpublished = {\url{https://meetobi.com/}},
  year = {2024},
  note = {Accessed: 2024-09-11}
}

@misc{bostondynamics2024,
  author = {{Boston Dynamics}},
  title = {Spot - The Agile Mobile Robot},
  howpublished = {\url{https://bostondynamics.com/products/spot/}},
  year = {2024},
  note = {Accessed: 2024-09-11}
}

@misc{naturallanguagedecomposition,
      title={Natural Language Decomposition and Interpretation of Complex Utterances}, 
      author={Harsh Jhamtani and Hao Fang and Patrick Xia and Eran Levy and Jacob Andreas and Ben Van Durme},
      year={2024},
      eprint={2305.08677},
      archivePrefix={arXiv},
      primaryClass={cs.CL},
      url={https://arxiv.org/abs/2305.08677}, 
}

@article{Asurveyofmultiagent,
title = {A survey of multi-agent Human–Robot Interaction systems},
journal = {Robotics and Autonomous Systems},
volume = {161},
pages = {104335},
year = {2023},
issn = {0921-8890},
doi = {https://doi.org/10.1016/j.robot.2022.104335},
url = {https://www.sciencedirect.com/science/article/pii/S092188902200224X},
author = {Abhinav Dahiya and Alexander M. Aroyo and Kerstin Dautenhahn and Stephen L. Smith},
}

@article{groundedlanguage,
title = {Grounded language interpretation of robotic commands through structured learning},
journal = {Artificial Intelligence},
volume = {278},
pages = {103181},
year = {2020},
issn = {0004-3702},
doi = {https://doi.org/10.1016/j.artint.2019.103181},
url = {https://www.sciencedirect.com/science/article/pii/S0004370218302935},
author = {Andrea Vanzo and Danilo Croce and Emanuele Bastianelli and Roberto Basili and Daniele Nardi},
}

@Article{AReviewofNaturalLanguage,
AUTHOR = {Liu, Rui and Guo, Yibei and Jin, Runxiang and Zhang, Xiaoli},
TITLE = {A Review of Natural-Language-Instructed Robot Execution Systems},
JOURNAL = {AI},
VOLUME = {5},
YEAR = {2024},
NUMBER = {3},
PAGES = {948--989},
URL = {https://www.mdpi.com/2673-2688/5/3/48},
ISSN = {2673-2688},
DOI = {10.3390/ai5030048}
}

@inproceedings{zhang2022smart,
  author    = {Zhang, T. and Zhao, D. and Yang, J. and Wang, S. and Liu, H.},
  title     = {A Smart Home Based on Multi-heterogeneous Robots and Sensor Networks for Elderly Care},
  booktitle = {Intelligent Robotics and Applications. ICIRA 2022},
  editor    = {Liu, H. and others},
  series    = {Lecture Notes in Computer Science},
  volume    = {13455},
  year      = {2022},
  publisher = {Springer, Cham},
  doi       = {10.1007/978-3-031-13844-7_10}
}

@article{luperto2023movecare,
  title={Integrating Social Assistive Robots, IoT, Virtual Communities and Smart Objects to Assist at-Home Independently Living Elders: the MoveCare Project},
  author={Luperto, M. and Monroy, J. and Renoux, J. and others},
  journal={International Journal of Social Robotics},
  volume={15},
  pages={517--545},
  year={2023},
  publisher={Springer},
  doi={10.1007/s12369-021-00843-0}
}

@article{UnifiedLearning,
author = {Mehta, Shaunak A. and Losey, Dylan P.},
title = {Unified Learning from Demonstrations, Corrections, and Preferences during Physical Human–Robot Interaction},
year = {2024},
issue_date = {September 2024},
publisher = {Association for Computing Machinery},
address = {New York, NY, USA},
volume = {13},
number = {3},
url = {https://doi.org/10.1145/3623384},
doi = {10.1145/3623384},
journal = {J. Hum.-Robot Interact.},
month = {aug},
articleno = {39},
numpages = {25}
}

@InProceedings{Few-ShotPreferenc,
  title = 	 {Few-Shot Preference Learning for Human-in-the-Loop RL},
  author =       {III, Donald Joseph Hejna and Sadigh, Dorsa},
  booktitle = 	 {Proceedings of The 6th Conference on Robot Learning},
  pages = 	 {2014--2025},
  year = 	 {2023},
  editor = 	 {Liu, Karen and Kulic, Dana and Ichnowski, Jeff},
  volume = 	 {205},
  series = 	 {Proceedings of Machine Learning Research},
  month = 	 {14--18 Dec},
  publisher =    {PMLR},
  url = 	 {https://proceedings.mlr.press/v205/iii23a.html},
}

@ARTICLE{Human-AwareRobot,
  author={Cheng, Yujiao and Sun, Liting and Tomizuka, Masayoshi},
  journal={IEEE Robotics and Automation Letters}, 
  title={Human-Aware Robot Task Planning Based on a Hierarchical Task Model}, 
  year={2021},
  volume={6},
  number={2},
  pages={1136-1143},
  doi={10.1109/LRA.2021.3056370}}

@article{Safetybounds,
title = {Safety bounds in human robot interaction: A survey},
journal = {Safety Science},
volume = {127},
pages = {104667},
year = {2020},
issn = {0925-7535},
doi = {https://doi.org/10.1016/j.ssci.2020.104667},
url = {https://www.sciencedirect.com/science/article/pii/S0925753520300643},
author = {Angeliki Zacharaki and Ioannis Kostavelis and Antonios Gasteratos and Ioannis Dokas}
}

@article{Perceivedsafety,
title = {Perceived safety in physical human–robot interaction—A survey},
journal = {Robotics and Autonomous Systems},
volume = {151},
pages = {104047},
year = {2022},
issn = {0921-8890},
doi = {https://doi.org/10.1016/j.robot.2022.104047},
url = {https://www.sciencedirect.com/science/article/pii/S0921889022000173},
author = {Matteo Rubagotti and Inara Tusseyeva and Sara Baltabayeva and Danna Summers and Anara Sandygulova},
}

@article{Improvinguserspecifications,
author = {Nils Wilde and Alexandru Blidaru and Stephen L Smith and Dana Kulić},
title ={Improving user specifications for robot behavior through active preference learning: Framework and evaluation},

journal = {The International Journal of Robotics Research},
volume = {39},
number = {6},
pages = {651-667},
year = {2020},
doi = {10.1177/0278364920910802},

URL = { 
    
        https://doi.org/10.1177/0278364920910802
    
    

},
eprint = { 
    
        https://doi.org/10.1177/0278364920910802
    
    

}
}

@INPROCEEDINGS{6251680,
  author={Hayashi, Kotaro and Sakamoto, Daisuke and Kanda, Takayuki and Shiomi, Masahiro and Koizumi, Satoshi and Ishiguro, Hiroshi and Ogasawara, Tsukasa and Hagita, Norihiro},
  booktitle={2007 2nd ACM/IEEE International Conference on Human-Robot Interaction (HRI)}, 
  title={Humanoid robots as a passive-social medium - a field experiment at a train station}, 
  year={2007},
  volume={},
  number={},
  pages={137-144},
  doi={10.1145/1228716.1228735}}

@INPROCEEDINGS{1573609,
  author={Hayashi, K. and Kanda, T. and Miyashita, T. and Ishiguro, H. and Hagita, N.},
  booktitle={5th IEEE-RAS International Conference on Humanoid Robots, 2005.}, 
  title={Robot Manzai - robots' conversation as a passive social medium}, 
  year={2005},
  volume={},
  number={},
  pages={456-462},
  doi={10.1109/ICHR.2005.1573609}}

@inproceedings{classmeta,
author = {Liu, Ziyi and Zhu, Zhengzhe and Zhu, Lijun and Jiang, Enze and Hu, Xiyun and Peppler, Kylie A and Ramani, Karthik},
title = {ClassMeta: Designing Interactive Virtual Classmate to Promote VR Classroom Participation},
year = {2024},
isbn = {9798400703300},
publisher = {Association for Computing Machinery},
address = {New York, NY, USA},
url = {https://doi.org/10.1145/3613904.3642947},
doi = {10.1145/3613904.3642947},
booktitle = {Proceedings of the CHI Conference on Human Factors in Computing Systems},
articleno = {659},
numpages = {17},
location = {Honolulu, HI, USA},
series = {CHI '24}
}

@ARTICLE{valley,
  author={Mori, Masahiro and MacDorman, Karl F. and Kageki, Norri},
  journal={IEEE Robotics \& Automation Magazine}, 
  title={The Uncanny Valley [From the Field]}, 
  year={2012},
  volume={19},
  number={2},
  pages={98-100},
  doi={10.1109/MRA.2012.2192811}}

@article{Unaldi_Yalcin_Elci_2023, title={An IoT-based Smart Home Application with Barrier-Free Stairs for Disabled/Elderly People}, volume={29}, url={https://eejournal.ktu.lt/index.php/elt/article/view/30731}, DOI={10.5755/j02.eie.30731}, abstractNote={&lt;p&gt;Home automation based on the Internet of Things (IoT) includes various components such as lighting, security, and remote control. Smart Home (SH) components should be varied and customised according to the user’s specific needs. Therefore, in this study, a SH system needed by users with specials is designed and implemented on a model. The NodeMCU microcontroller with ESP8266 Wi-Fi module, Radio Frequency Identification (RFID) tags, temperature-humidity, motion detection, gas and moisture sensors and several actuators are used to build the system. Thanks to If This Then That (IFTTT), Google Assistant, and Blynk, the SH components can be managed remotely and with voice commands via a user-friendly Android-based mobile interface. The multiple control system and the ability to control the home components with different methods make this study comprehensive. In particular, the barrier-free stairs design has added innovation to the SH system for disabled people. Thus, the accessibility, security, and comfort requirements of disabled people are met, and their quality of life is improved to live independently.&lt;/p&gt;}, number={1}, journal={Elektronika ir Elektrotechnika}, author={Unaldi, Sibel and Yalcin, Nesibe and Elci, Enes}, year={2023}, month={Feb.}, pages={15-20} }

@INPROCEEDINGS{MergingPhysicalandSocial,
  author={Nguyen, Phuong D.H. and Bottarel, Fabrizio and Pattacini, Ugo and Hoffmann, Matej and Natale, Lorenzo and Metta, Giorgio},
  booktitle={2018 IEEE-RAS 18th International Conference on Humanoid Robots (Humanoids)}, 
  title={Merging Physical and Social Interaction for Effective Human-Robot Collaboration}, 
  year={2018},
  volume={},
  number={},
  pages={1-9},
  doi={10.1109/HUMANOIDS.2018.8625030}}

@inproceedings{AgreeingtoInteract,
author = {Sasabuchi, Kazuhiro and Ikeuchi, Katsushi and Inaba, Masayuki},
title = {Agreeing to Interact: Understanding Interaction as Human-Robot Goal Conflicts},
year = {2018},
isbn = {9781450356152},
publisher = {Association for Computing Machinery},
address = {New York, NY, USA},
url = {https://doi.org/10.1145/3173386.3173390},
doi = {10.1145/3173386.3173390},
booktitle = {Companion of the 2018 ACM/IEEE International Conference on Human-Robot Interaction},
pages = {21–28},
numpages = {8},
location = {Chicago, IL, USA},
series = {HRI '18}
}

@inproceedings{DeepLearningandSentiment,
author = {Atzeni, Mattia and Reforgiato Recupero, Diego},
title = {Deep Learning and Sentiment Analysis for Human-Robot Interaction},
year = {2018},
isbn = {978-3-319-98191-8},
publisher = {Springer-Verlag},
address = {Berlin, Heidelberg},
url = {https://doi.org/10.1007/978-3-319-98192-5_3},
doi = {10.1007/978-3-319-98192-5_3},
booktitle = {The Semantic Web: ESWC 2018 Satellite Events: ESWC 2018 Satellite Events, Heraklion, Crete, Greece, June 3-7, 2018, Revised Selected Papers},
pages = {14–18},
numpages = {5},
location = {Heraklion, Greece}
}

@article{Multi-domainsentiment,
title = {Multi-domain sentiment analysis with mimicked and polarized word embeddings for human–robot interaction},
journal = {Future Generation Computer Systems},
volume = {110},
pages = {984-999},
year = {2020},
issn = {0167-739X},
doi = {https://doi.org/10.1016/j.future.2019.10.012},
url = {https://www.sciencedirect.com/science/article/pii/S0167739X19309719},
author = {Mattia Atzeni and Diego Reforgiato Recupero}
}

@article{humanrobotcollaboration,
title = {Human-robot collaboration: A multilevel and integrated leadership framework},
journal = {The Leadership Quarterly},
volume = {33},
number = {1},
pages = {101594},
year = {2022},
note = {The Leadership Quarterly Yearly Review (LQYR) for 2022},
issn = {1048-9843},
doi = {https://doi.org/10.1016/j.leaqua.2021.101594},
url = {https://www.sciencedirect.com/science/article/pii/S1048984321000990},
author = {Chou-Yu Tsai and Jason D. Marshall and Anwesha Choudhury and Andra Serban and YoYo {Tsung-Yu Hou} and Malte F. Jung and Shelley D. Dionne and Francis J. Yammarino}
}

@article{explainablegoal,
author = {Sado, Fatai and Loo, Chu Kiong and Liew, Wei Shiung and Kerzel, Matthias and Wermter, Stefan},
title = {Explainable Goal-driven Agents and Robots - A Comprehensive Review},
year = {2023},
issue_date = {October 2023},
publisher = {Association for Computing Machinery},
address = {New York, NY, USA},
volume = {55},
number = {10},
issn = {0360-0300},
url = {https://doi.org/10.1145/3564240},
doi = {10.1145/3564240},
journal = {ACM Comput. Surv.},
month = {feb},
articleno = {211},
numpages = {41}
}

@inproceedings{augmentedmath,
author = {Chulpongsatorn, Neil and Lunding, Mille Skovhus and Soni, Nishan and Suzuki, Ryo},
title = {Augmented Math: Authoring AR-Based Explorable Explanations by Augmenting Static Math Textbooks},
year = {2023},
isbn = {9798400701320},
publisher = {Association for Computing Machinery},
address = {New York, NY, USA},
url = {https://doi.org/10.1145/3586183.3606827},
doi = {10.1145/3586183.3606827},
booktitle = {Proceedings of the 36th Annual ACM Symposium on User Interface Software and Technology},
articleno = {92},
numpages = {16},
location = {San Francisco, CA, USA},
series = {UIST '23}
}

@misc{MobileViT,
      title={MobileViT: Light-weight, General-purpose, and Mobile-friendly Vision Transformer}, 
      author={Sachin Mehta and Mohammad Rastegari},
      year={2022},
      eprint={2110.02178},
      archivePrefix={arXiv},
      primaryClass={cs.CV},
      url={https://arxiv.org/abs/2110.02178}, 
}

@INPROCEEDINGS{MobiFace,
  author={Duong, Chi Nhan and Quach, Kha Gia and Jalata, Ibsa and Le, Ngan and Luu, Khoa},
  booktitle={2019 IEEE 10th International Conference on Biometrics Theory, Applications and Systems (BTAS)}, 
  title={MobiFace: A Lightweight Deep Learning Face Recognition on Mobile Devices}, 
  year={2019},
  volume={},
  number={},
  pages={1-6},
  doi={10.1109/BTAS46853.2019.9185981}}

@article{designandvalidation,
title = {Design and validation of a light-weight reasoning system to support remote health monitoring applications},
journal = {Engineering Applications of Artificial Intelligence},
volume = {41},
pages = {232-248},
year = {2015},
issn = {0952-1976},
doi = {https://doi.org/10.1016/j.engappai.2015.01.019},
url = {https://www.sciencedirect.com/science/article/pii/S0952197615000305},
author = {Aniello Minutolo and Massimo Esposito and Giuseppe {De Pietro}}
}

@ARTICLE{MobiHisNet,
  author={Kumar, Abhinav and Sharma, Anshul and Bharti, Vandana and Singh, Amit Kumar and Singh, Sanjay Kumar and Saxena, Sonal},
  journal={IEEE Internet of Things Journal}, 
  title={MobiHisNet: A Lightweight CNN in Mobile Edge Computing for Histopathological Image Classification}, 
  year={2021},
  volume={8},
  number={24},
  pages={17778-17789},
  doi={10.1109/JIOT.2021.3119520}}

\appendix

\end{document}